\documentclass[usenatbib]{mnras}

\usepackage[T1]{fontenc}
\usepackage{ae,aecompl}

\usepackage{graphicx}	
\usepackage{amsmath}	
\usepackage{amssymb}	
\usepackage{subfigure}
\usepackage{cancel}

\title[Stellar populations in green valley galaxies]{A detailed look at the stellar populations in green valley galaxies}

\author[Angthopo, Ferreras \& Silk]
{James Angthopo$^{1}$, Ignacio Ferreras$^{1,2,3}$\thanks{E-mail: i.ferreras@ucl.ac.uk},
Joseph Silk$^{4,5,6}$
\\
$^1$ Mullard Space Science Laboratory, University College London,
Holmbury St Mary, Dorking, Surrey RH5 6NT, UK\\
$^2$ Department of Physics and Astronomy, University College London,
Gower Street, London WC1E 6BT, UK\\
$^3$ Instituto de Astrof{\'i}sica de Canarias, Calle V{\'i}a L{\'a}ctea s/n,
E38205, La Laguna, Tenerife, Spain\\
$^4$ Institut d'Astrophysique de Paris (UMR 7095: CNRS \& UPMC), 98 bis Bd Arago, F-75014 Paris, France\\
$^5$ Sub-department of Astrophysics, University of Oxford, Keble Road, Oxford OX1 3RH, UK\\
$^6$ Department of Physics and Astronomy, The Johns Hopkins University Homewood Campus, Baltimore, MD 21218, USA
}

\date{MNRAS. Accepted 2020 May 4. Received 2020 March 29; in original form 2019 August 28}

\pubyear{2020}

\hypersetup{%
    ,urlcolor=black
    ,citecolor=black
    ,linkcolor=black
}

\begin{document}
\label{firstpage}
\pagerange{\pageref{firstpage}--\pageref{lastpage}}
\maketitle

\begin{abstract}
The green valley (GV) represents an important transitional
state from actively star-forming galaxies to passively evolving
systems. Its traditional definition, based on colour, rests on a
number of assumptions that can be subject to non-trivial
systematics. In \citet{Ang:19}, we proposed a new
definition of the GV based on the 4000\AA\ break strength.  In this
paper, we explore in detail the properties of the underlying stellar
populations by use of $\sim$230 thousand high-quality spectra from the
Sloan Digital Sky Survey (SDSS), contrasting our results with a
traditional approach via dust-corrected colours. We explore high
quality stacked SDSS spectra, and find a population trend that
suggests a substantial difference between low- and high-mass galaxies,
with the former featuring younger populations with star formation
quenching, and the latter showing older (post-quenching) populations
that include rejuvenation events. Subtle but measurable
differences are found between a colour-based approach and our
definition, especially as our selection of GV galaxies produces 
a cleaner ``stratification'' of the GV, with more homogeneous
population properties within sections of the GV.
Our definition based on  4000\AA\ break strength gives a
clean representation of the transition to quiescence, easily
measurable in the upcoming and future spectroscopic surveys.
\end{abstract}




\section{Introduction} \label{sec:Intro}

Our understanding of galaxy formation and evolution has greatly
advanced over the past decades, especially thanks to high quality
all-sky galaxy surveys. However, challenges lie ahead, due to the
complex mixture of physical processes, the different timescales
involved, and the inherent degeneracies in the interpretation of the
observations. To tackle these hurdles, we have to combine numerical
hydrodynamical simulations based on physical equations, with large
galaxy surveys, where the large data sets allow us to carefully select
samples, control systematics, and follow a general statistical
approach. These surveys often combine photometric and spectroscopic
data, enabling the discovery of fundamental relations such as a
conspicuous bimodality \citep[see, e.g.,][]{Strateva:01}.  The bimodal
behaviour represents a clear divide regarding star formation activity
(or stellar population age) with a strong dependence with stellar mass
and cosmic age \citep[see, e.g.,][]{Bell:04}.  This bimodality can be
shown on various diagrams spanned by parameters such as the
colour-magnitude \citep{2009Gr}, star formation rate (SFR) - mass
\citep{Schiminovich_2007, 2013ZUE}, UVJ bi-colour diagram
\citep{2009Will}, or colour-mass \citep{ 2018GAMAGV,Schawinski2014},
to name a few.  The two contrasting galaxy populations have been
termed the red sequence (hereafter RS), and the blue cloud (hereafter
BC) in colour-mass diagrams.  Inherent to this bimodal nature is the
region separating BC from RS, commonly known as the green valley
\citep[hereafter GV,][]{Salim:14}.

By construction, the GV can be considered a region overlapping a
number of possible evolutionary pathways between BC and RS, and
various scenarios have been postulated to understand the observed
distribution of galaxies \citep{2007Faber}. It is generally accepted
that galaxies on the GV are undergoing quenching of their star
formation activity (roughly ``ascending'' on a colour vs stellar mass
diagram, where the colour index increases upwards), or undergoing
rejuvenation events \citep{Thomas:10} from recent infall of gas --
therefore tracing a ``descending'' track on this diagram.  Multiple
survey-based studies have been carried out to assess the properties of
GV galaxies. \citet{Martin:07} combined NUV and optical fluxes to
define and study the GV, finding a significant increase in the
fraction of AGN, with respect to BC or RS galaxies. The connection of
AGN activity with quenching was further supported by the transition
phases found in the population analysis of SDSS early-type galaxies
made by \citet{KSchaw:07}. However, a morphological analysis of GV
galaxies revealed a complex mixture \citep{Schawinski2014}, leading to
a scenario where, at least, two channels are present, with short
quenching timescales ($\sim$100\,Myr) being involved in the evolution
of early-type galaxies, and longer timescales ($\sim$2-3\,Gyr) in
late-type systems \citep[see
  also,][]{Smethurst:15,NC:18}. \citet{Phillipps:19} use {\sc MAGPHYS}
to extract star formation histories from GAMA GV galaxies to derive
transition times $\sim$2--4\,Gyr, with no specific signal that the
quenching takes place faster than an otherwise decaying
rate. Moreover, the transition processes appear to affect mostly the
disk component, favouring secular disk fading \citep{2018GAMAGV}, and
environment unsurprisingly affecting the appearance of the GV
\citep{GVEnv}. It is worth mentioning here that the presence of
merger-like morphologies is not favoured in GV galaxies
\citep{Mendez:11}. Also note that the GV transition times appear to
proceed faster at high redshift, following the standard downsizing
scenario \citep[see, e.g.,][for a study at z$\sim$0.8]{Goncalves:12}.

Regarding intrinsic colour distributions, it is found that GV galaxies
present blue outer regions, so that the recent star formation
responsible for their being GV galaxies may be caused by rejuvenation
events from the infall of gas clouds or gas-rich smaller galaxies
\citep{Thilker:10,SR:10,Fang:12}. Alternatively, one can consider the
quenching of star formation moving outwards from the centre due to gas
depletion \citep{2018GAMAGV2}.

On the theoretical side, state-of-the-art numerical simulations of
galaxy formation such as EAGLE are able to quantify quenching
timescales \citep{Wright:19} and the physical processes associated
with quenching \citep{QuenchMethods}.  Through these simulations, they
find that the quenching timescale is dependent on galaxy mass and
environment. The simulations suggest that low-mass galaxies,
M$_{\star}<10^{9.6}$\,M$_{\odot}$, feature quenching timescales
$\gtrsim$3\,Gyr, and intermediate mass galaxies,
$10^{9.7}$\,M$_{\odot}<$M$_{\star}< 10^{10.3}$\,M$_{\odot}$, have the
longest quenching timescales, whereas the most massive galaxies,
M$_{\star}>10^{10.3}$\,M$_{\odot}$ are estimated to have the shortest
quenching timescales, $\tau\lesssim$2\,Gyr. Moreover, quenching is
faster in satellites, with respect to centrals \citep[however
  see,][]{Trayford:16}.  Note though, that alternative selection
methods, such as using sub-millimeter fluxes, give rise to different
morphologies of the same region \citep[e.g. ``green
  mountain'',][]{2018GMount}, reflecting the complexity of the
interpretation of this transition region, and the potential biases
caused by the specific details of the selection.

The wide range of timescales found suggests a mixture of evolutionary
channels.  At low mass, quenching may be mostly due to ram pressure
stripping or stellar feedback, depending on whether the galaxies are
satellites or centrals. Intermediate mass galaxies, with the highest
quenching timescales, are thought to undergo radio-mode AGN and/or
stellar feedback. Finally, at the massive end, major mergers that
include strong AGN activity seem to be the main cause of quenching
\citep{2006EarlyAGN}. Moreover, halo mass may also provide a valid
mode of quenching in galaxies hosted by halos above a critical mass
M$_{\rm crit}\sim10^{12}$\,M$_{\odot}$ \citep{QuenchMethods}.
However, to further complicate our attempt at understanding the nature
of GV galaxies, they are also likely to move from RS to GV or even
into the BC.  This can occur through events such as wet mergers, where
a quiescent gas-poor galaxy merges with a star-forming gas-rich
galaxy, so that the surplus of gas causes rejuvenation
\citep{Thomas:10}. Events such as accretion of gas may also cause
rejuvenation.

Due to the importance of the GV as a transition phase that can
constrain the underlying physical processes, it is essential that we
find an effective and robust definition of the GV, that is easily
implemented in theoretical models of galaxy formation. Although many
intriguing results have been observed using the colour-based
definition of the GV, they may suffer from systematics caused by the
adopted dust correction. The standard procedure \citep[see,
  e.g.,][]{Brinchmann:04} compares a number of photometric and
spectroscopic observables with a set of population synthesis models
where a dust prescription is applied, assuming an extinction law fixed
to constraints from Milky Way stars \citep{1989Card} or from starburst
galaxies \citep{Calz:00}. This method appears to work quite well
\citep{Ang:19}, however the results may be prone to uncontrolled
systematics, especially given the observed correlation between the
parameters that describe the {\sl effective attenuation} by dust in
star-forming galaxies \citep[see,
  e.g.][]{KC:13,Tress:18,Salim:18,Narayanan:18}.

This paper focuses on an analysis of the new selection of GV galaxies
using the 4000\AA\ break strength as proposed by \citet[][hereafter
  A19]{Ang:19}.  The 4000\AA\ break is a highly sensitive age
indicator defined over a relatively narrow spectral region
\citep[250\AA\ using the definition of][]{Balogh:99}, to avoid a large
effect from dust, but wide enough to be measured with relatively low
S/N spectra, including low spectral resolution, as, e.g. with slitless
grism data \citep{Hathi:09} or medium band filters \citep{AHC:13}.
Therefore, the break can be accurately measured at moderately low
spectral resolution, opening its use to present and future medium
spectral resolution surveys that use slitless grism spectroscopy, such
as PEARS \citep{PEARS}; FIGS \citep{FIGS}; Euclid \citep{Euclid};
WFIRST \citep{WFIRST}; or imaging surveys that use medium band
passbands, such as ALHAMBRA \citep{ALHAMBRA}; SHARDS \citep{SHARDS};
JPAS \citep{JPAS}.  The definition of the GV with the 4000\AA\ break
is expected to provide a more direct representation of the age
distribution, with substantially weaker contamination from dust.

Hereafter, we refer to the GV selection via the 4000\AA\ strength as
``D4k sel''.  We include a comparative study with a colour-based
selection, via $^{0.1}(g-r)$ -- hereafter defined as ``$^{0.1}(g-r)$
sel''. Note the colour is K-corrected to z=0.1, that represents the
typical redshift of our SDSS-based sample. Section \ref{sec:Data}
outlines the survey used, as well as the creation of the spectral
stacks used in the analysis of GV galaxies.  Section
\ref{sec:LSAnalysis} explores the stacks by use of line strengths and
simple stellar population (SSP) models. Section \ref{sec:CompAnalysis}
analyses the properties of GV galaxies using spectral fitting of
composite populations. In Section \ref{sec:Discussion} we discuss the
main results, and Section \ref{sec:Conc} finishes with a summary.

\section{Sample selection}  \label{sec:Data}
\subsection{Spectroscopic data}

This paper continues the work presented in A19, based on a sample of
spectroscopic data from the classic Sloan Digital Sky Survey
\citep[SDSS,][]{SDSS:00}. We select the sample and retrieve the
spectra from Data Release~14 \citep{SDSS:DR14}. SDSS is a full-sky
survey that includes spectroscopic observations of galaxies with
Petrosian $r$-band magnitude in the range 14.5$<r_{\rm AB}<$17.7. Our
selection criteria identifies targets with relatively high
signal-to-noise ratio, snMedian$\_$r$>$10.  We stack the spectra
within carefully defined regions on the plane that defines GV galaxies
(see below).

Our GV is defined on a plane spanned by a stellar population parameter
(i.e. either the 4000\,\AA\ break strength or a more standard
broadband colour) along with velocity dispersion ($\sigma$), as
measured in the SDSS fibre. $\sigma$ is preferred with respect to
stellar mass, as it is directly measurable in good quality spectra,
being free of the systematics associated with stellar mass estimates
(such as the derivation of the total flux, or the model-dependent
constraint of the stellar mass to light ratio). Furthermore, velocity
dispersion is the observable that more strongly correlates with
stellar population properties \citep[see,
  e.g.,][]{Ber:03,SAMIGrad}. The use of spectra with the imposed
threshold in S/N guarantees a robust estimate of $\sigma$.  Our sample
is constrained in redshift between z=0.05 and 0.1, (median $0.077\pm
0.013$) resulting in $\sim$228,000 galaxies with high quality spectra.
We bin the sample according to velocity dispersion from
$\sigma$=70\,km/s to 250\,km/s, with bin spacing of $30\,$km/s.

The SDSS spectra cover the wavelength range  3800--9200\,\AA,
with variable resolution, from R=1,500 at $\lambda$=3,800\,\AA,
increasing to R=2,500 at $\lambda$=9,000\,\AA\ \citep{smee:13}. The spectra are
dereddend with respect to foreground extinction, adopting the standard
Milky Way dust law \citep{1989Card}.  The correction required was
determined by the extinction parameter $A_g$, quoted in the SDSS catalogues,
evaluated in the $g$ band. Once the foreground dust correction
is applied, the spectra are brought to the rest frame, and the individual
estimates of the 4000\AA\ break are measured, adopting the following
definition \citep{Balogh:99}:
\begin{equation}
  {\rm D}_n(4000)=\frac{\langle\Phi^R\rangle}{\langle\Phi^B\rangle}, \;
  {\rm where}\;
  \langle\Phi^i\rangle\equiv\frac{1}{(\lambda_2^i-\lambda_1^i)}
    \int_{\lambda_1^i}^{\lambda_2^i}\Phi(\lambda)d\lambda
\end{equation}
and $(\lambda_1^B,\lambda_2^B,\lambda_1^R,\lambda_2^R)=(3850,3950,4000,4100)$\AA.
Note that our definition contrasts with the traditional approach, originally
defined by \citet{GB:83} that integrate $\Phi(\nu)$ in wavelength space.
We believe our definition has an easier interpretation, as the flux ratio between
two spectral intervals that straddle the 4000\AA\ break. These two
definitions can be compared via a rescaling with a 
constant factor $\sim$1.08 between the old and the new definition, irrespective of
the properties of the underlying populations.

In this paper, as well as in A19, we contrast
the newly defined GV with the traditional one based on colours from
broadband photometry. We adopt $^{0.1}(g-r)$ as the reference colour, i.e.
all galaxies are measured at a fiducial redshift z=0.1.
The colours are taken from the flux ratios as measured within the 3\,arcsec
fibres of the SDSS classic spectrograph -- to be consistent with the
analysis of the observed spectra -- and the K-correction needed to
bring them to the fiducial redshift is measured directly from the spectra,
following the standard approach \citep[see, e.g.][]{Kcorr}, adopting a
vanilla-standard cosmology ($\Omega_m=0.3$; $\Omega_\Lambda=0.7$,
$H=70$\,km\,s$^{-1}$\,Mpc$^{-1}$). As reference, our K-correction in $(g-r)$
stays below 0.14\,mag (below 0.085\,mag in 90\% of the sample),
and the median correction applied is $+0.04\pm0.03$\,mag. 
The colour is also corrected for
{\sl intrinsic} dust absorption, following the dust parameters of
\citet{2003KaSM} and adopting the \citet{Calz:00} attenuation law.
Note we use $A_V$ as baseline for the correction, where we find, at
z=0.1, $A_g^{0.1}=1.16A_V^{0.1}$ and $A_r^{0.1}=0.88A_V^{0.1}$.
We refer the reader to fig.~2 in A19 for a comparison between
dust-corrected and uncorrected colours on the selection plane.
Note the classic SDSS 3\,arcsec diameter fibres map the 
central region of galaxies, $\sim$3-5\,kpc. However the interpertation
of our results are robust (see Appendix~\ref{app:aperture} for an
analysis of aperture effects). 

\begin{figure*}
    \centering
    \includegraphics[width=0.47\linewidth]{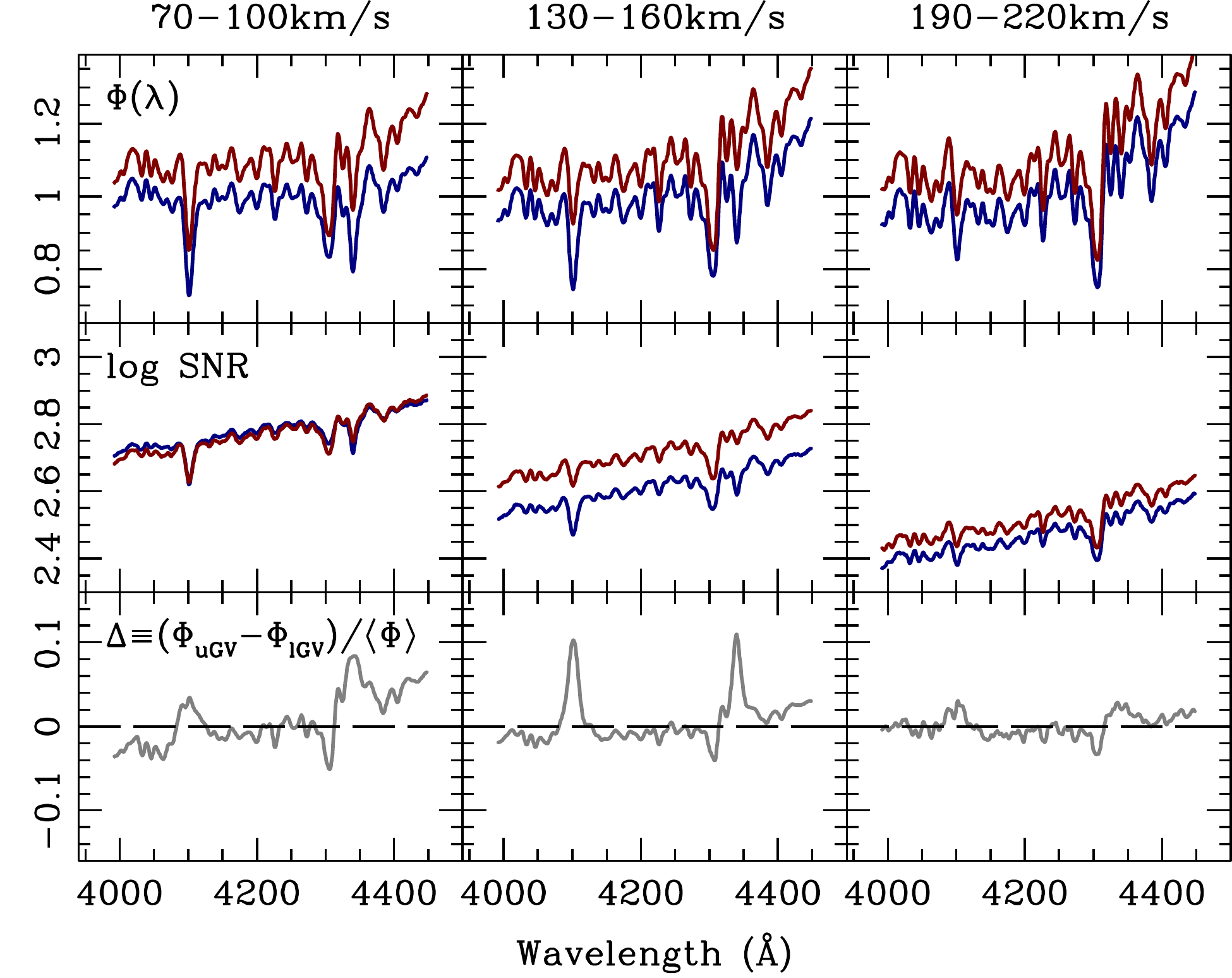}
    \includegraphics[width=0.47\linewidth]{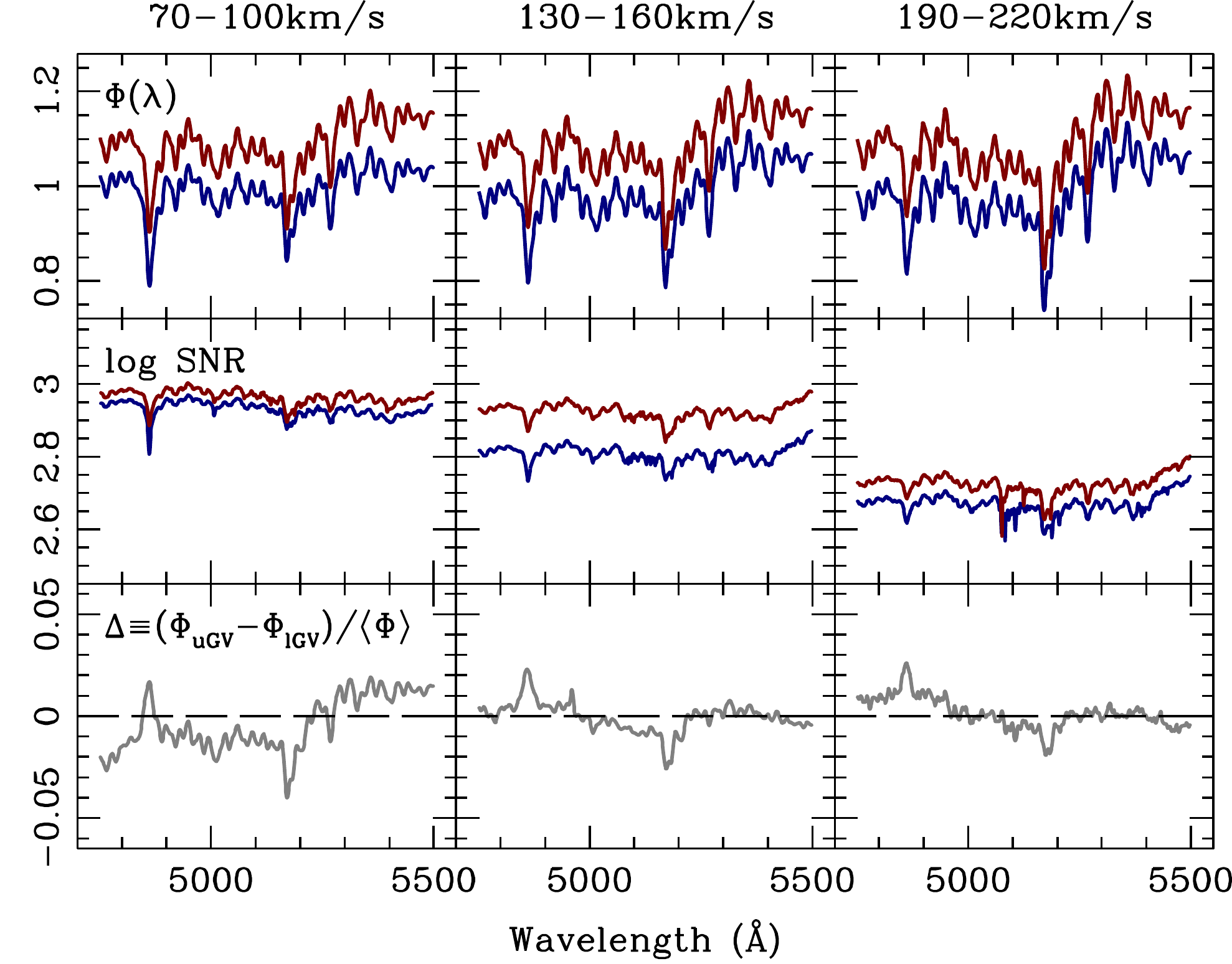}
    \caption{A comparison of the stacks is shown here in two important
      spectral windows: the neighbourhood of the H$\gamma$ and
      H$\delta$ absorption lines (left panels) and the region around
      the H$\beta$ and Mg+Fe complex (right panels). From top to
      bottom, the panels show the fluxes of the lGV (in blue) and the
      uGV (in red), with a small arbitrary vertical offset to avoid crowding;
      the signal to noise ratio on a logarithmic scale; and the
      relative flux difference between uGV and lGV spectra. The
      top labels show the velocity dispersion in each case. 
      All stacks have been convolved to a common velocity
      dispersion equivalent of 235\,km/s.}
    \label{fig:stacks}
\end{figure*}

\subsection{GV definition}
\label{ssec:GVDef}

We give here a brief description of the selection of GV galaxies, as
proposed by A19, and refer interested readers to that paper for more
details. Each galaxy is identified by two parameters: velocity
dispersion ($\sigma$) and an indicator of stellar age (hereafter
$\pi$), choosing either the 4000\AA\ break strength, D$_n$(4000), or
the dust-corrected colour, $^{0.1}(g-r)$. In addition, the sample is
cross-matched with the JHU/MPA catalogue \citep{2003KaSM} from which
we retrieve the BPT classification of nebular emission lines
\citep{BPT1981}. This enabled us to split the sample into
unclassified/quiescent (BPT=$-$1), star-forming (BPT=1,2), composite
(BPT=3), Seyfert (BPT=4) and LINER (BPT=5) galaxies. We do not consider here
those spectra classified as composite (BPT=3), for a cleaner
classification of the three regions. Moreover, the definition of the
GV, as shown below, only relies on quiescent and star-forming galaxies.
In  each velocity
dispersion bin, we fit separately the distribution of
star-forming (SF) and quiescent (Q) galaxies adopting 
a Gaussian distribution with respect to $\pi$, namely:
\begin{equation}
  {\cal P}_k(\pi; \sigma) \equiv \frac{1}{s_k(\sigma)\sqrt{2\pi}}
  e^{-\frac{1}{2}\left[\frac{\pi-\mu_k(\sigma)}{s_k(\sigma)}\right]^2},
  \label{eq:Prob_Gen}
\end{equation}
where $\mu_k(\sigma)$ and $s_k(\sigma)$ are, respectively, the mean and the standard
deviation of the distribution of $\pi$ corresponding to
galaxies in the velocity dispersion bin given by $\sigma$.
We now propose the ansatz that these Gaussian distributions be
interpreted as the probability distribution function (PDF) of
blue cloud galaxies (for $k$=SF) and red sequence galaxies (for $k$=Q).
Once the PDFs are defined for the BC and RS, 
the green valley subset is assumed to follow a probability distribution
function given by:
\begin{equation}
  {\cal P}_{\rm GV}(\pi; \sigma) \equiv \frac{1}{s_{\rm GV}(\sigma)\sqrt{2\pi}}
  e^{-\frac{1}{2}\left[\frac{\pi-\mu_{\rm GV}(\sigma)}{s_{\rm GV}(\sigma)}\right]^2},
\end{equation}
where the width of the Gaussian is chosen 
\begin{equation}
  s_{\rm GV}(\sigma) = \frac{1}{2}s_{\rm Q}(\sigma),
\end{equation}
and the mean is given by
\begin{equation}
  {\cal P}_{\rm SF}[\mu_{\rm SF}(\sigma); \sigma] =
  {\cal P}_{\rm Q}[\mu_{\rm Q}(\sigma); \sigma].
\end{equation}
We emphasize these constraints are purely empirical and defined ad hoc.
The constraint on the mean implies that at the peak of the GV PDF,
a BC galaxy and a RS galaxy are indistinguishable from a
probabilistic point of view. The constraint on the width
ensures that the GV does not include large fractions of galaxies
in the BC or RS regions.
Note that this method is performed independently within each velocity dispersion
bin. The actual selection of GV galaxies follows a Monte Carlo sampling
method. For each galaxy within a given velocity dispersion bin, a
uniform random deviate ($r$) is obtained between 0 and 1, and the galaxy
is accepted into the GV subset if ${\cal P}_{\rm GV}[\pi; \sigma]>r$.
The probability distribution functions
obtained for the six velocity dispersion bins can be found in fig.~1 of A19.
The GV set is further split into an upper- (uGV), middle- (mGV) and lower- (lGV)
green valley, defined by the terciles of the distribution of $\pi$ in the
GV sample within each velocity dispersion bin.
Table~\ref{tab:fractions}, in the appendix,
shows the number of galaxies in the uGV and lGV within each velocity dispersion bin and
the fraction of galaxies according to their BPT flag. For reference, we
include in the table the results when colour is not corrected with respect
to intrinsic dust attenuation.

\subsection{Spectral Stacking}

The subsamples of GV galaxies are then used to produce high quality
stacked spectra. High S/N is needed for a robust analysis of the
stellar population content. However, more importantly, our motivation
to stack the spectra is to average out galaxy-to-galaxy variations,
leading to a set of ``super-spectra'' for which the variations between
different regions of the selection plane are only caused by the
transitional mechanisms that give rise to the BC/GV/RS distribution.

We follow the standard procedure for the stacking of the SDSS spectra
\citep[see, e.g.][]{IF:13}. The stacking was carried on spectra that
were de-reddened and brought to a rest-frame wavelength in the air
system, applying a normalisation according to the median flux in the
rest-frame interval 5000--5500\AA. The process implies resampling the
flux within each spectral pixel following a linear split between
adjacent pixels according to the amount of overlap between the
original pixel and the sampled pixel. Each resulting stack is then
corrected for nebular emission by performing spectral fitting with the
{\sc STARLIGHT} code \citep{STARLIGHT05}. In the stacking procedure,
we exclude Seyfert AGN (BPT=4) and Composite systems (BPT=3), since
prominent AGN luminosity contaminates the continuum, affecting the
colours and the 4000\AA\ break strength, parameters used in the
definition of the GV. The residuals with respect to the best fit
spectrum are then used to fit Gaussian profiles within the standard
emission regions, which are then removed from the stacks (see A19 and
\citealt{FLB:13} for details).  In order to compare the spectral
features across the wide range of velocity dispersion, we convolve all
stacks with a Gaussian kernel to produce a velocity dispersion of
235\,km/s in all cases. Fig.~\ref{fig:stacks} compares the stacked
data of uGV and lGV galaxies in two important regions: the interval
around the age-sensitive Balmer indices H$\gamma$ and H$\delta$ (left
panels) and the region covering the metallicity-sensitive indices Mgb
and $\langle$Fe$\rangle$ (right panels). From top to bottom, we show
the stacked spectra; the S/N; and the difference between uGV and lGV
spectra, in three velocity dispersion bins, as labelled.

\begin{figure*}
    \centering
    \includegraphics[width=0.45\linewidth]{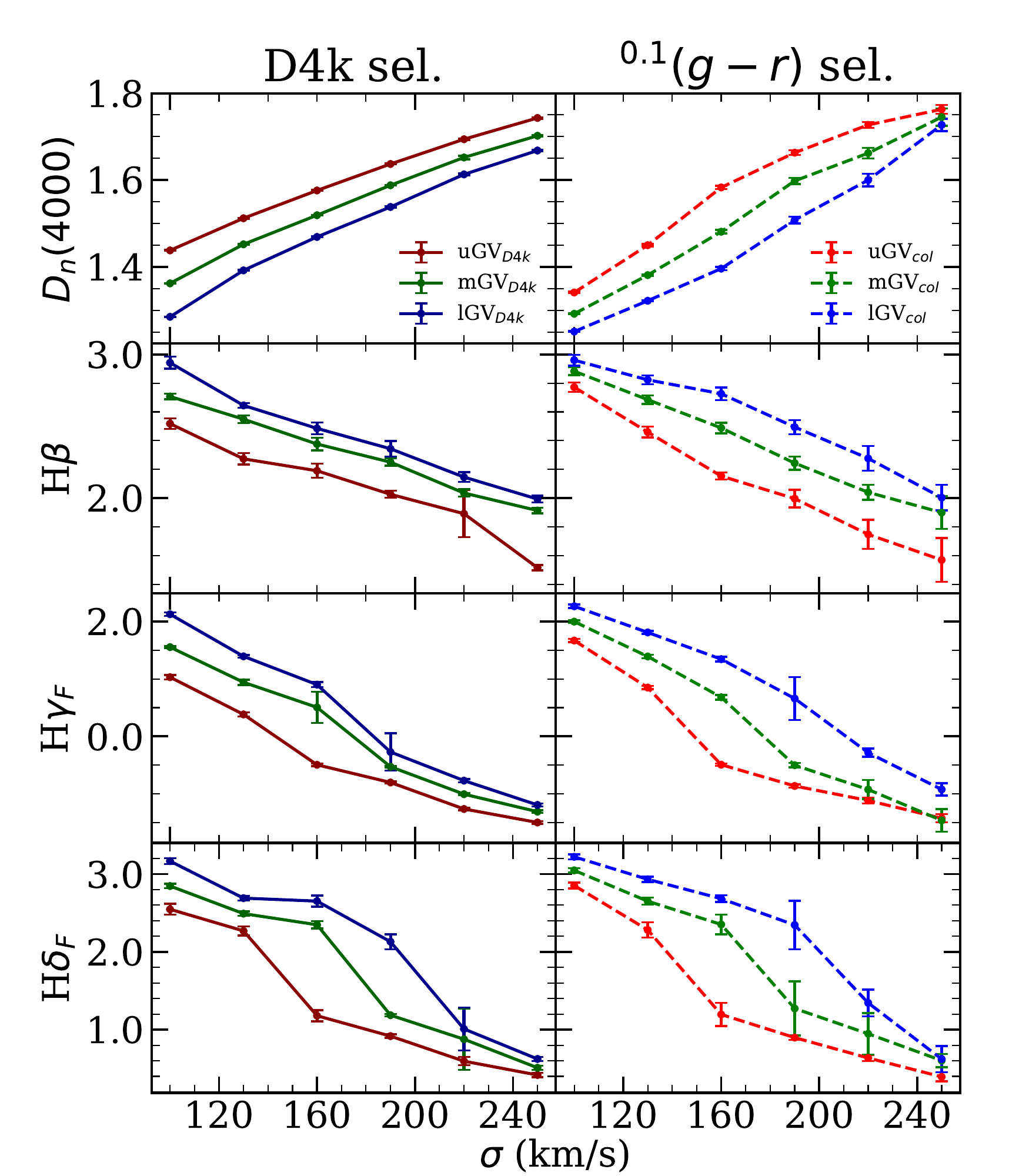}
    \includegraphics[width=0.45\linewidth]{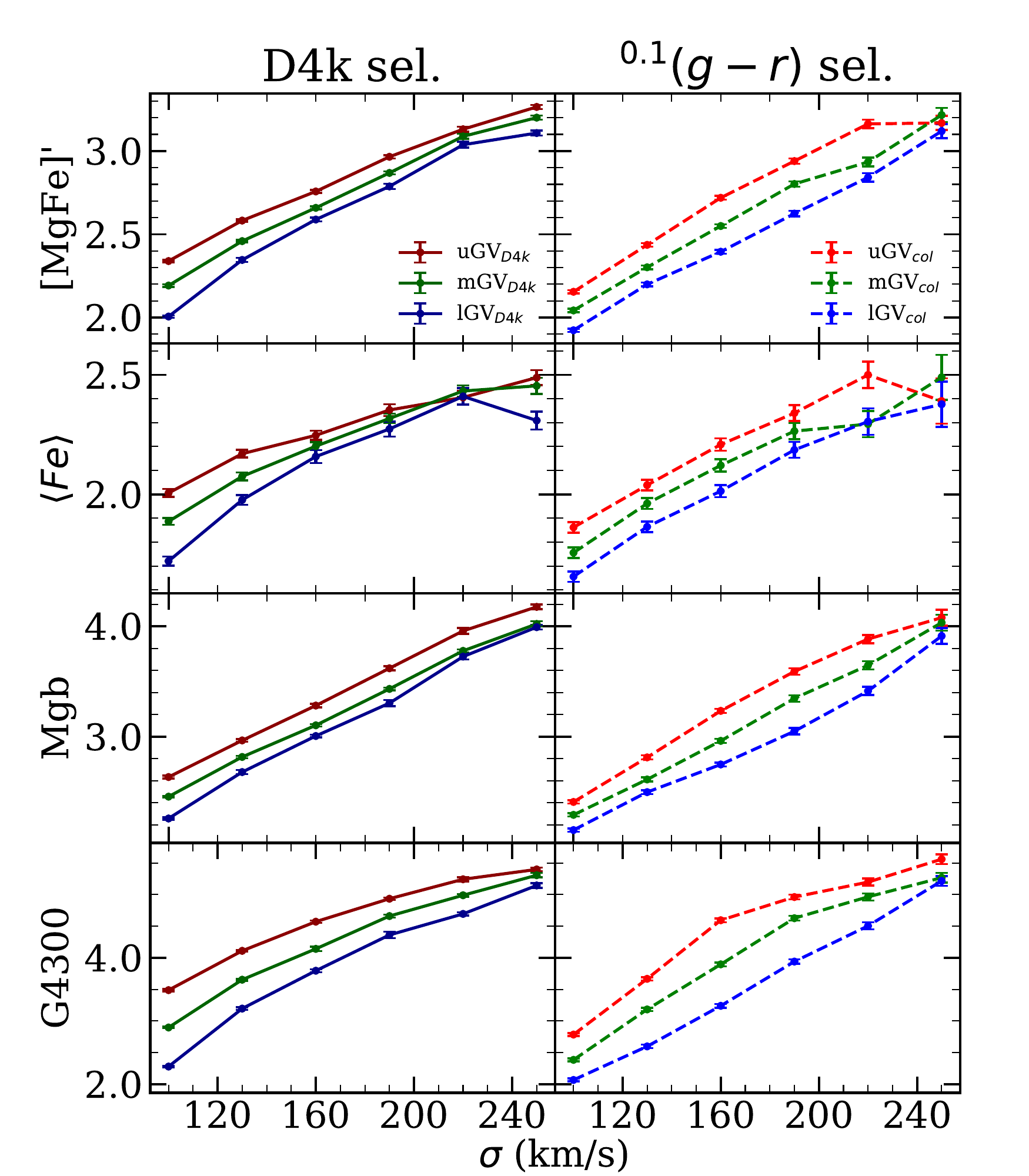}
    \caption{Comparison between line strengths of GV galaxies
      selected by 4000\AA\ break strength (D4k) or dust-corrected
      $^{0.1}(g-r)$ colour (col). Each panel shows the results
      for the upper, middle and lower GVs, as labelled. {\sl Left:} From top
      to bottom, we show the age-sensitive indices
      $D_n(4000)$, $H\beta$, $H\gamma_F$ and
      $H\delta_F$. {\sl Right:} metallicity-sensitive indices (from top to bottom),
      [MgFe]$^\prime$, $\langle$Fe$\rangle$, Mgb, and the age-sensitive index G4300.
      All measurements are given as equivalent widths, in \AA, except for $D_n(4000)$,
    which is a dimensionless ratio.}
    \label{fig:LineS}
\end{figure*}

\subsection{Uncertainty of the stacked spectra}
\label{sec:Uncert}

One of the key constraints in our sample selection is the signal to
noise ratio of individual spectra (snMedian\_r$>$10) in order to avoid
stacking large numbers of noisy data.  This constraint results in
stacked spectra with a very high S/N (see Fig.~\ref{fig:stacks}) when
computed with the standard statistical noise carried from the
individual data.  Such high values of S/N lead to underestimated
uncertainties on the derived parameters, as the higher values of the
best-fit $\chi^2$ reflect the shortcomings of population synthesis
models at this level of detail. Therefore to ensure our results have
more realistic uncertainties, we adapt the noise level including two
estimates -- added in quadrature -- that take into account additional
sources of uncertainty in the stacking procedure.  (1) We create Monte
Carlo realisations of each stack by using the uncertainty of
individual fluxes. We carry out the same analysis for these stacks as
our original set, therefore giving us a more robust statistical
uncertainty of the derived parameters. (2) For each velocity
dispersion bin, we bootstrap the subsample, selecting, at random, only
$60 \%$ of the galaxies.  We carry out the same process as for the
original stacks, therefore incorporating the systematical uncertainty
in our error bars caused by the sample selection. Bins comprising
fewer galaxies are expected to carry a larger uncertainty, accounting
for sample selection systematics.

Additional systematics may be expected, inherently to the methodology 
adopted here.  One such systematic relates to the use of SSP
models \citep{BC03, MIU:12}, that carry their own systematic
uncertainties that depend upon the stellar library, isochrones and
initial mass function (IMF) chosen. Furthermore, the use of {\sc
  STARLIGHT} to perform spectral fitting will carry additional
uncertainties.  One way to mitigate this systematic would involve
comparisons among independent spectral fitting algorithms, such as
pPXF \citep{2004pPXF} or FIREFLY \citep{Firefly:17}, beyond the scope
of this paper.  Another source of uncertainty arises from the emission
line correction that we apply to the Balmer absorption lines. However,
our use of a battery of emission line diagnostics, and the comparison
with spectral fitting -- that mask out such regions --  mitigates 
this potential systematic.

\section{Line strength analysis} \label{sec:LSAnalysis}

Our first approach to the analysis of the underlying stellar
populations in GV galaxies focuses on the observed absorption line strengths. We
select a battery of standard indices from the Lick system: $H\beta$,
Mgb, Fe5270, Fe5335, G4300 \citep{1998Trager}, the higher order
Balmer indices $H\gamma_F$ and $H\delta_F$ \citep{1997Worthey}, as well as the
$D_n(4000)$ \citep{Balogh:99} index already used for the definition of the
GV. We combine the iron abundance indices into an average
$\langle$Fe$\rangle$\,$\equiv$(Fe5270+Fe5335)/2, and also measure the
standard index
[MgFe]$^\prime\equiv\sqrt{{\rm Mgb}(0.72\times{\rm Fe5270}+ 0.28\times{\rm Fe5335})}$ 
\citep{2003DThomas}. These indices can be split into two categories --
age- and metallicity-sensitive. The Balmer indices, $D_n(4000)$ and G4300
are usually considered age-sensitive, while the others are
metallicity-sensitive. However any and all spectral indices mentioned
here (and, unfortunately, elsewhere) suffer from the age-metallicity
degeneracy \citep[e.g.,][]{1994AgeZDeg, FCS:99}.

\subsection{General trends}

This  approach, solely based on the observed line strengths, is meant to assess in a 
model-independent way the variations between different regions
across the GV.  Fig.~\ref{fig:LineS} shows the 
line strengths measured in the stacks, plotted with respect to
velocity dispersion. We note these measurements are taken from the stacks
that are smoothed to a common velocity dispersion of 235\,km/s.
Within each subfigure, the left (right) panels correspond to a selection
of GV galaxies based on 4000\AA\ break strength (dust-corrected $^{0.1}(g-r)$ colour).
Each panel shows independently the trends in the upper, middle and lower GV,
as labelled, including an error bar obtained from bootstrapping the sample. The 
$D_n(4000)$ index increases strongly with velocity dispersion in both cases and in all
three regions of the GV, in agreement with previous studies
\citep[see, e.g.,][]{2003KaSM,2009Gr}. However, at low velocity dispersion, the
D4k selection produces overall higher values of the break, along with
a wider range, towards lower values of the Balmer indices,
consistently supporting the hypothesis that the D4k selection
produces a GV with older and more homogeneous populations, especially at the low-mass end.
At the high mass end, both selection criteria give rather similar results, although
the lGV set defined by D4k appears to give slightly lower values than the
colour-based lGV. However, this behaviour is not
paralleled by the Balmer lines, so the mapping into population trends becomes
less trivial. Nevertheless, the strong trends found in all age-sensitive
indices with respect to velocity dispersion give robust confirmation of
the well-known age-mass relation \citep[see, e.g.][]{AnnaGallD4k}.

\begin{figure*}
  \centering
  \includegraphics[width=0.8\linewidth]{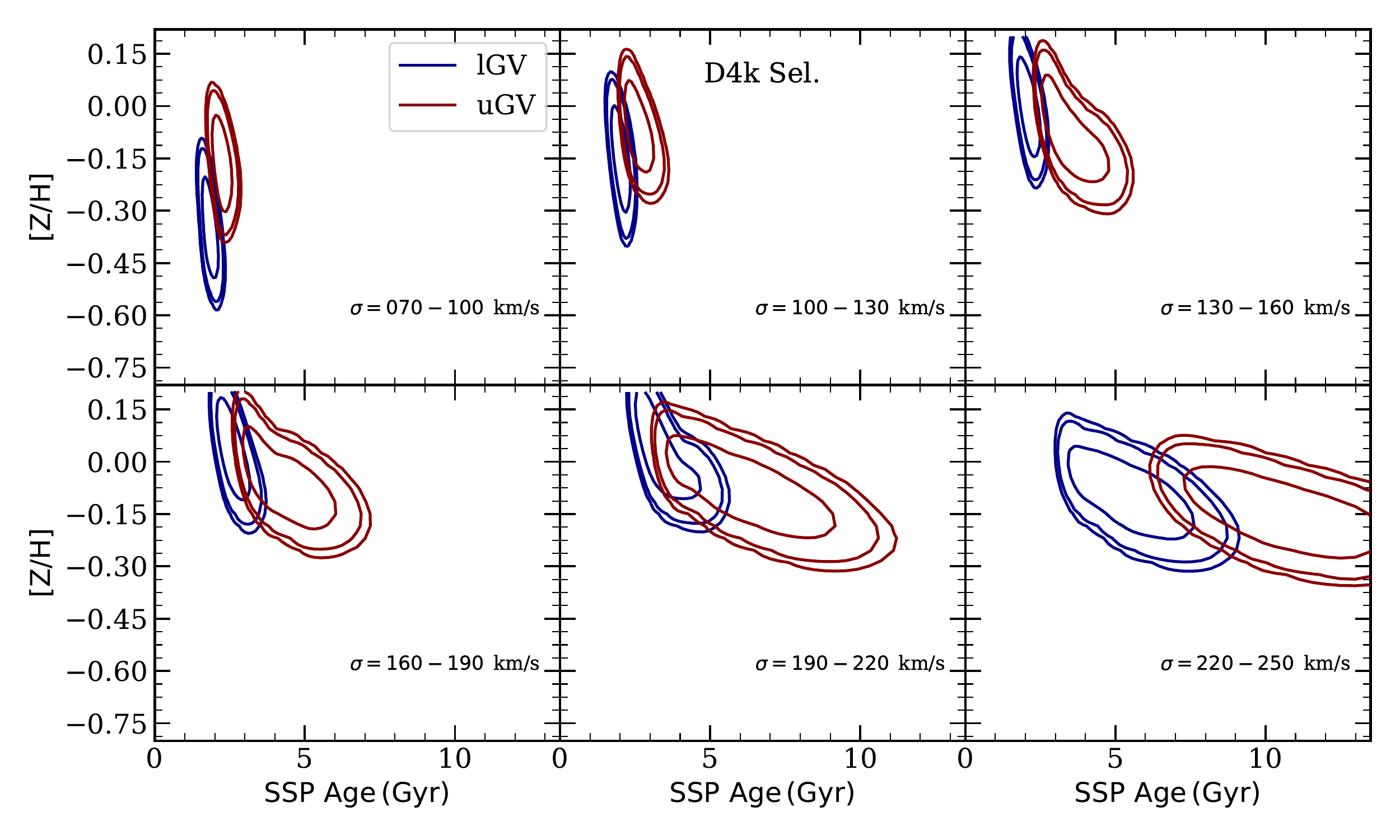}
  \caption{The likelihood derived from the line strength analysis of
    SSP-equivalent age and metallicity is shown with contour plots at
    the equivalent 1, 2 and 3\,$\sigma$ levels. These stacks
    correspond to the selection of GV galaxies based on 4000\AA\ break
    strength. The blue (red) contours represent lGV (uGV) stacks.}
  \label{fig:2Dcont}
\end{figure*}

Both selection methods
unsurprisingly produce the lowest (i.e. youngest) break strengths and
highest (i.e. youngest) Balmer indices in the lGV at the low-mass end,
whereas at high velocity dispersion, the differences between lGV and
uGV are minor. We should keep in mind that the age-sensitive indices
presented in Fig.~\ref{fig:LineS} behave differently with
respect to the age distribution, with the Balmer indices being more
sensitive to recent star formation episodes compared to $D_n(4000)$
\citep{D4000ref}. Regardless of the GV selection method, as we traverse from lGV
to uGV, we find a decrease in the line strengths of all Balmer
indices, thus indicating a smooth transition from the younger lGV to
the predominantly quenched, passively evolving uGV. It is worth noting that the colour-based
selection produces overall higher Balmer indices, possibly implying the
presence of younger (but dustier) galaxies, with respect to the D4k selection.
Even though the colour selection includes an intrinsic
dust correction, biases may appear as the correction is 
prone to systematics regarding the derivation of the dust corrections as well
as variations of the dust attenuation law caused by the diverse range of
dust geometry and chemical composition
in galaxies \cite[see, e.g.][]{Tress:18,Narayanan:18}. Moreover, note that
the K-correction applied to bring the observed colour within a
fiducial value (z=0.1 in our case) may also introduce an additional systematic:
the median K-correction applied is 0.04\,mag but there is an obvious
dependence with redshift -- with the correction trivially vanishing at
z=0.1 -- and a more subtle trend with the intrinsic colour -- with the
K-correction being larger in the redder galaxies. The difference in this correction with respect
to the intrinsic colour can be as high as $\sim$0.1\,mag, thus comparable with
the interval that defines the GV. Such behaviour can
introduce a correlation in the selection of the GV. In contrast, the
4000\AA\ break selection depends neither on dust correction (see Appendix~\ref{app:dust})
nor on the K-correction (as the index is directly measured on the
rest-frame spectra).

The panels on the right of Fig.~\ref{fig:LineS} present the results for
metal-sensitive indices along with G4300.
The index G4300 increases with velocity
dispersion, roughly following a very similar trend as $D_n(4000)$.
Note that this index separates more smoothly the lGV, mGV and uGV,
confirming a strong correlation with the 4000\AA\ break.
For [MgFe]$^\prime$,
$\langle$Fe$\rangle$ and Mgb, we find the expected positive
correlation with increasing velocity dispersion
\citep[e.g.,][]{2008LineBimod}. The line strength [MgFe]$^\prime$ -- that can
be approximately considered a total metallicity indicator -- shows
that, in general, the D4k selection features slightly more metal rich
populations compared to the colour selection, in all GV regions. By
comparing $\langle$Fe$\rangle$ and Mgb, we find both produce similar
trends. As we move up the GV, from lGV to uGV, the data show 
increasing metallicity. It is worth mentioning here that the dependence 
of these indices on age -- due to the age-metallicity degeneracy -- is
such that higher values of the index could also be explained by {\sl older}
ages. Section~\ref{sec:CompAnalysis} is devoted to a comparison of the
stacked spectra with population synthesis models via spectral fitting, to
be able to break such degeneracies. Moreover, we show below (Section~\ref{sec:SSP})
an analysis of stellar ages based on simple stellar populations.

Therefore, Fig.~\ref{fig:LineS} shows a subtle but measurable difference with respect
to the GV selection method (D4k vs colour). These differences are especially substantial
at low velocity dispersion, where the contribution from dusty, star-forming galaxies
may introduce a larger systematic on the dust correction needed when using
a colour selection.

\subsection{Simple Stellar Population (SSP) properties}
\label{sec:SSP}

In A19 we provided an estimate of the SSP-equivalent ages of GV stacks
restricted to the subset of quiescent galaxies. Here, we include
star-forming and LINER-like AGN, to produce stacks that give a more
comprehensive description of the average properties of GV galaxies.
We also include more information about the SSP model fitting procedure
-- which is identical to the one presented in A19 -- and extend the
analysis, including a simple measurement of non-solar abundance
ratios. The interpretation of the stacked spectra is done via
a comparison of a selected set of age- and metallicity-sensitive line
strengths, between the observed measurements and the values obtained
from stellar population synthesis models.

\subsubsection{Age and metallicity}

In this section, we produce easy-to-interpret `SSP-equivalent' ages,
instead of a more complex combination of populations, left to
Section~\ref{sec:CompAnalysis}. SSP-derived ages should be interpreted
as a luminosity averaged age, as if the whole stellar content of the
galaxy were formed in a single burst.  An alternative definition --
based on composite age distributions following a predefined functional
form of the star formation rate -- can be prone to biases due to the
specific form adopted. We show both estimates of age in this paper to
be able to assess the actual variations in the underlying populations
of GV galaxies.  We follow a Bayesian
approach, probing a large volume of SSPs from the MIUSCAT population
synthesis models \citep{MIU:12}, comparing the observed and the model
line strengths with a standard $\chi^2$ statistic:
\begin{equation}
  \chi^2 (t,Z)=\sum_i \left[\frac{\Delta_i(t,Z)}{\sigma_i}\right]^2,
  \label{eq:chi2}
\end{equation}
where $\Delta_i(t,Z)=O_i-M_i(t,Z)$ is the difference between the observed line strength
and the model prediction for the $i$th index, and $\sigma_i$ is the
corresponding uncertainty.
The grid of SSP models comprise 8,192 synthetic spectra. The stellar age
ranges from 0.1 to 13.5\,Gyr, in 128 logarithmically-spaced steps,
and metallicity ([Z/H]), varies from $-$2.0 to $+$0.2\,dex, with 64 steps.
We note the original models have a reduced set of metallicity steps (seven in total)
and we interpolate (bi)linearly for a given choice of (log) age and metallicity.

Since the signal-to-noise ratio of the data is very high (Fig.~\ref{fig:stacks}), we need to
apply offsets to the individual indices to account for potential
variations due to differences in the [Mg/Fe] abundance ratio of the
populations or, indeed, due to an extended age distribution. Our modus
operandi involves computing the best fit solution (i.e. giving the
minimum $\chi^2$) for a fiducial stack. This fiducial stack is chosen
as the one that gives the lowest value of $\chi^2$ for the best fit. We then
define the offsets for each line strength from this best fit solution,
and apply these offsets ($\delta_i$), such that
$\Delta_i(t,Z)=O_i-M_i(t,Z)-\delta_i$ in equation~\ref{eq:chi2} --
to {\sl all} the stacks in the sample. Given the large S/N of the spectra,
we also add in quadrature -- as a potential systematic error, and in
order to produce conservative error bars -- an additional amount
corresponding to 5\% of the measured line strength.  The resulting
$\chi^2$ distributions are bivariate functions of age and
metallicity. We fix the stellar initial mass function to \citet{Kroupa:01}.
Note that the results for
alternative choices of the IMF, such as Chabrier or Salpeter, give very
similar results, and that, within the range of velocity dispersion values
considered in this sample, no substantial variations of the IMF are
expected \citep[see, e.g.][]{IF:13}.

\begin{figure}
  \centering
  \includegraphics[width=85mm]{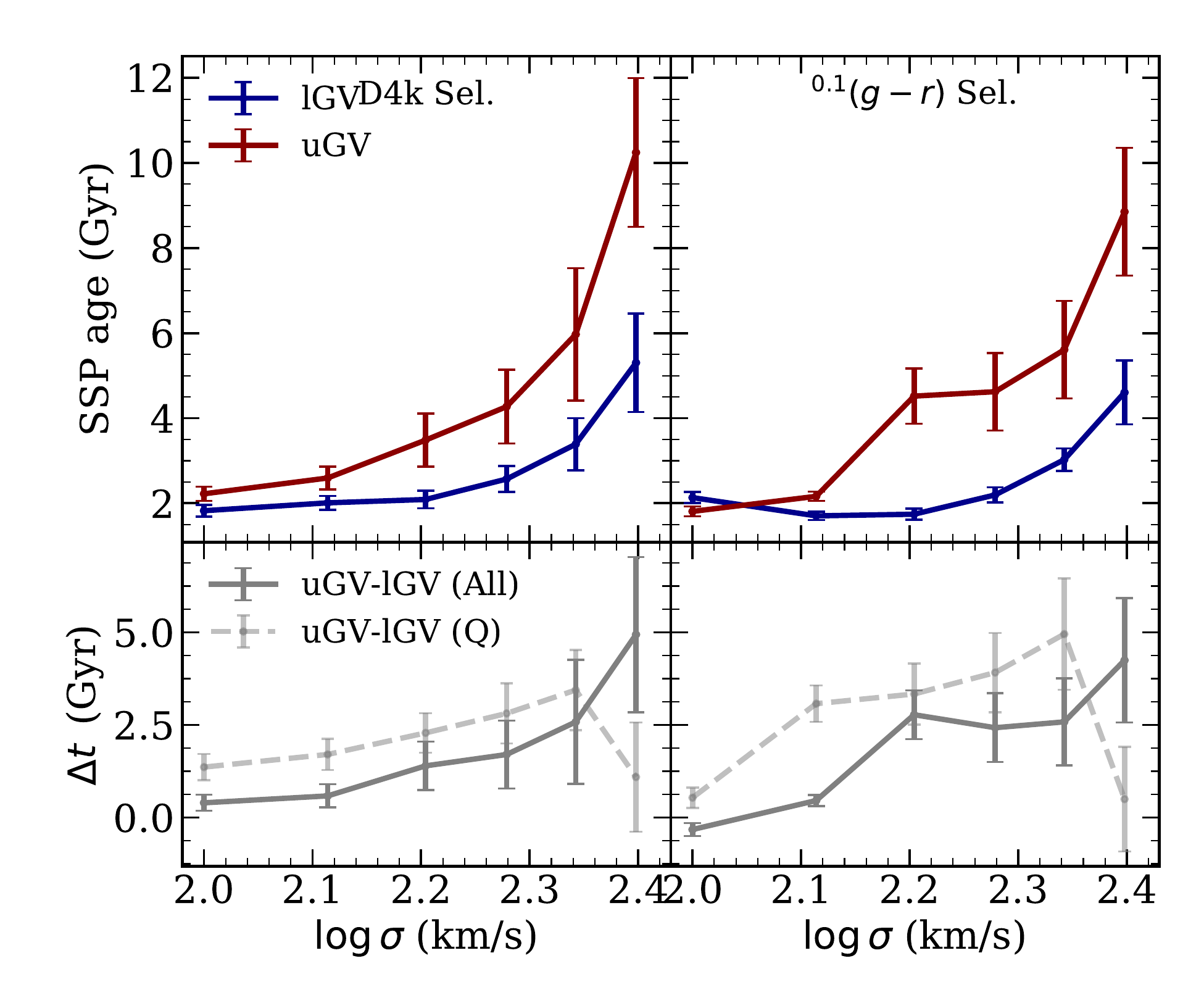}
  \caption{The SSP-equivalent age derived from line strengths is shown
    with respect to velocity dispersion in the lGV (blue) and uGV
    (red) subsets for the D4k- (left) and the colour-selected (right)
    GV. The bottom panels show the age difference between uGV and lGV
    stacks.  These are derived from the stacks presented in this
    paper, that comprise quiescent, star-forming and LINER AGN
    galaxies. The dashed grey lines in the bottom panels contrast the
    results with A19, where the stacks only involve quiescent GV
    galaxies.}
  \label{fig:SSPAge}
\end{figure}

Fig.~\ref{fig:2Dcont} shows the corresponding probability contours of
the bivariate likelihood ${\cal P}(t,Z)$ derived from the spectral
stacks of the uGV (red) and lGV (blue) galaxies, selected according to
4,000\AA\ break strength. The contours are slightly smoothed by a
Gaussian kernel, and shown at the equivalent 1, 2 and 3\,$\sigma$
confidence levels, with each panel representing a velocity dispersion
bin. Note the expected positive correlation between velocity
dispersion and either SSP-equivalent age or metallicity. We note that
the method is especially good for determining {\it relative}
variations in the stellar age, whereas metallicity is less well
constrained. The difference in age between lGV and uGV galaxies is
apparent. We marginalize over metallicity, producing the trends in
SSP-equivalent age shown in Fig.~\ref{fig:SSPAge}, with the error bars
given at the 1\,$\sigma$ level.  We show the uGV and lGV trends with
respect to velocity dispersion in the D4k-selected (left) and
colour-selected (right) GV. The bottom panels show the age difference
between the two. We stress that this paper focuses on spectral stacks
that include star-forming, quiescent and LINER-like galaxies. The
dashed grey lines in the bottom panels show the analysis when
restricting the stacks to quiescent galaxies, as shown in fig. 4 of
A19. Both sets of stacks feature a similar increasing trend with $\sigma$,  except at the 
highest velocity dispersion bin, where the quiescent sample shows a significant
decrease.

\subsubsection{[Mg/Fe]}

Overabundances in [Mg/Fe] are traditionally associated with short and
intense star formation episodes where the delayed Fe-rich contribution
from type Ia supernovae is not incorporated into stars \citep[see,
  e.g.,][]{Thomas:99}. Standard models based on a single or double
degenerate progenitor imply delays between 0.5 and 2\,Gyr \citep{MR:01}. Therefore,
populations with super-solar [Mg/Fe] are expected to have been formed 
over similar timescales. Therefore, an estimate of [Mg/Fe] provides
a stellar clock that has been exploited, for instance, to show that
massive early-type galaxies must have formed their central regions
within a dynamical timescale. Here we look for potential variations of
[Mg/Fe] in the stacked GV spectra, following the proxy adopted in 
\citet{FLB:13}. This proxy allows one to use standard, solar-scaled population
synthesis models to measure [Mg/Fe]. The procedure
involves constraining a grid of SSP models using two different sets of
line strengths, one involving age-sensitive indices along with 
an Mg-sensitive index (obtaining a
metallicity [Z$_{\rm Mg}$/H]; in this case we use Mgb) and another one involving
the same age-sensitive indices plus a Fe-sensitive
index (producing [Z$_{\rm Mg}$/H], we use here $\langle$Fe$\rangle$).
The difference, i.e. [Z$_{\rm Mg}$/Z$_{\rm Fe}$] is the adopted proxy for [Mg/Fe].

\begin{figure}
    \centering
    \includegraphics[width=75mm]{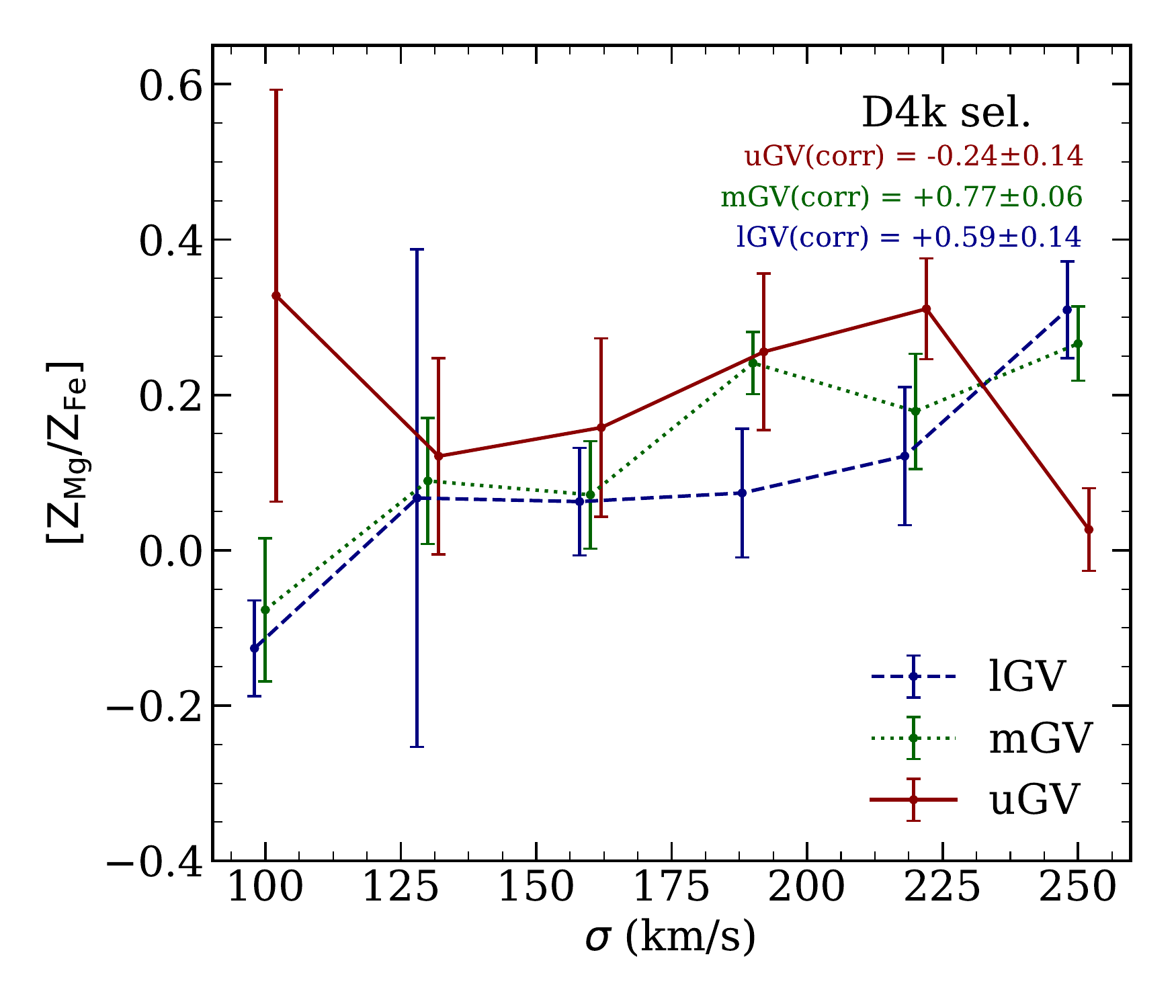}
    \includegraphics[width=75mm]{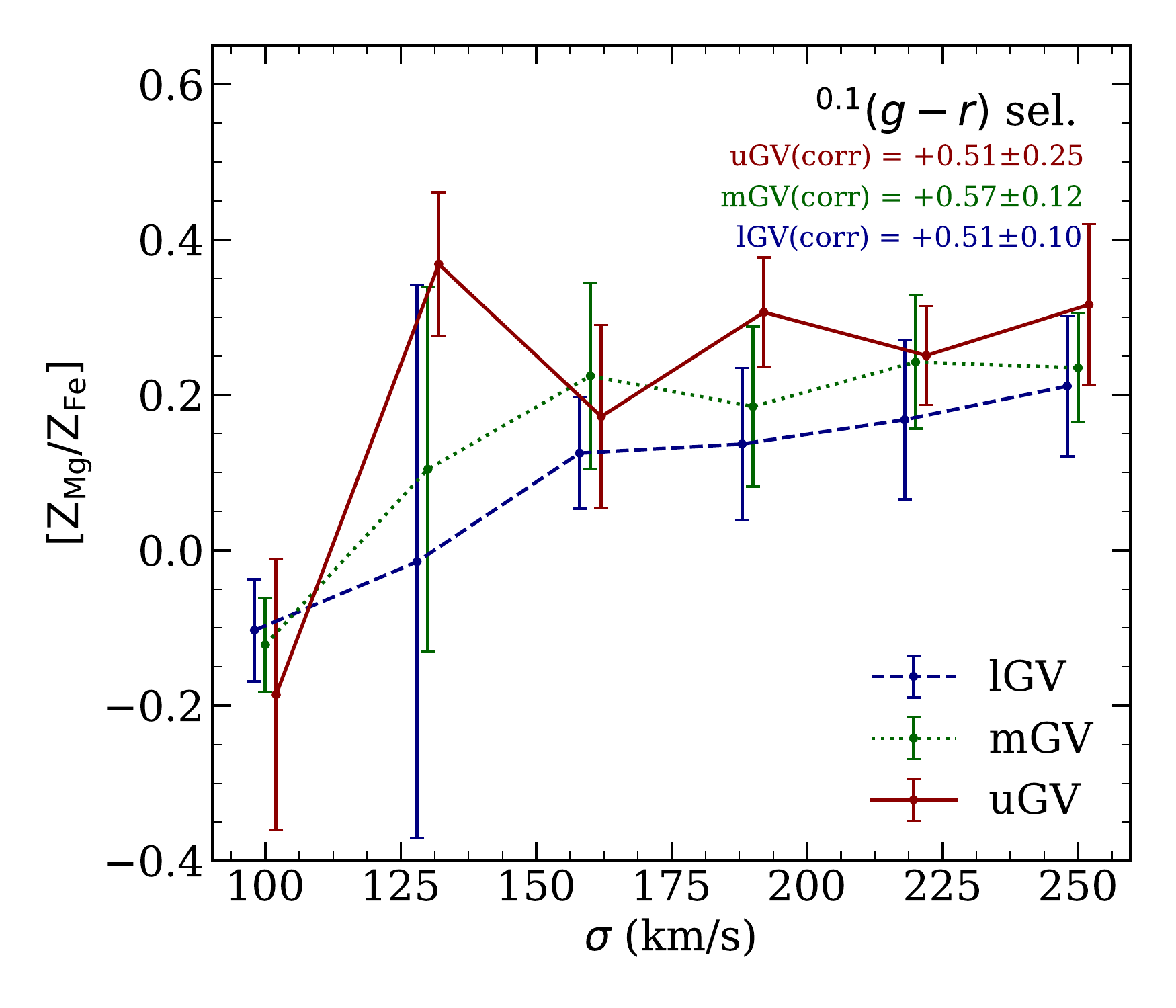}
    \caption{The relation between the abundance ratio proxy 
      $\mathrm{[Z_{Mg}/Z_{Fe}]}$ and velocity dispersion for 4000\AA\ break
      strength (top) and colour selection (bottom). The blue, green and
      red lines show the trends for lGV, mGV and uGV galaxies, respectively (see text for
      details about the derivation of this [Mg/Fe] proxy). The numbers in red, green, and
      blue show the Pearson correlation coefficient (pcc) between
      $\mathrm{[Z_{Mg}/Z_{Fe}]}$ and velocity dispersion in the
      uGV, mGV and lGV subsamples, respectively.}
    \label{fig:MgFe_Z}
\end{figure}

Fig.~\ref{fig:MgFe_Z} plots [Z$_{\rm Mg}$/Z$_{\rm Fe}$] against
velocity dispersion for both definitions of the GV, where the numbers
in the legend represent the Pearson correlation coefficient (pcc).
Our estimate of pcc along with its uncertainty involves 100 Monte
Carlo realisations. For each realisation, we remove at random one data
point and calculate the resulting pcc; this is done to mitigate the
effect of outliers. The pcc is quoted as the median of the
distribution and the uncertainty is one standard deviation. The data
points are shifted horizontally by an arbitrary amount to avoid
overcrowding. In the D4k selection (top panel), the uGV behaves
differently with respect to lGV and mGV, which behave in a similar
manner to each other. Both mGV (pcc=$+0.77\pm0.06$) and lGV
(pcc=$+0.59\pm0.14$) show a weak but significant correlation with
velocity dispersion, whereas the abundance ratio in uGV galaxies
(pcc=$-0.24\pm0.14$) seems rather constant.  Interestingly, the
colour-based selection (bottom panel) shows an increasing trend in all
subsamples of the GV, although [Z$_{\rm Mg}$/Z$_{\rm Fe}$] appears to
flatten in uGV galaxies with $\sigma\gtrsim$150\,km/s.  The trends
support the scenario of a more extended star formation history in
lower mass galaxies \citep[see, e.g.,][]{IGDR:11}.

\subsection{Difference regarding (nebular) activity} \label{sec:Diff_Stacks}

In A19, we restricted the stacking procedure to quiescent
galaxies. The motivation in that letter was to assess the transition
period of quenched galaxies across the GV, by considering systems
where star formation is already absent. In this paper, we consider,
instead, the general trends of GV galaxies, so that the different
contributions of quiescent, star-forming, and AGN systems
(see Table~\ref{tab:fractions}) manifests in those trends. Moreover,
as systems with nebular emission are expected to be significantly
affected by dust, we want to explore the differences between the D4k
selection (minimally affected by dust) and the colour selection (that
applies a dust correction, but may be affected by systematics from 
this correction). Therefore, a comparison between the quiescent
stacks of A19 and those that include star forming and
LINER-like emission, provides useful insight on the properties of
GV galaxies and the potential biases caused by the selection method.

\begin{figure*}
    \centering
    \includegraphics[width=\linewidth, height=70mm]{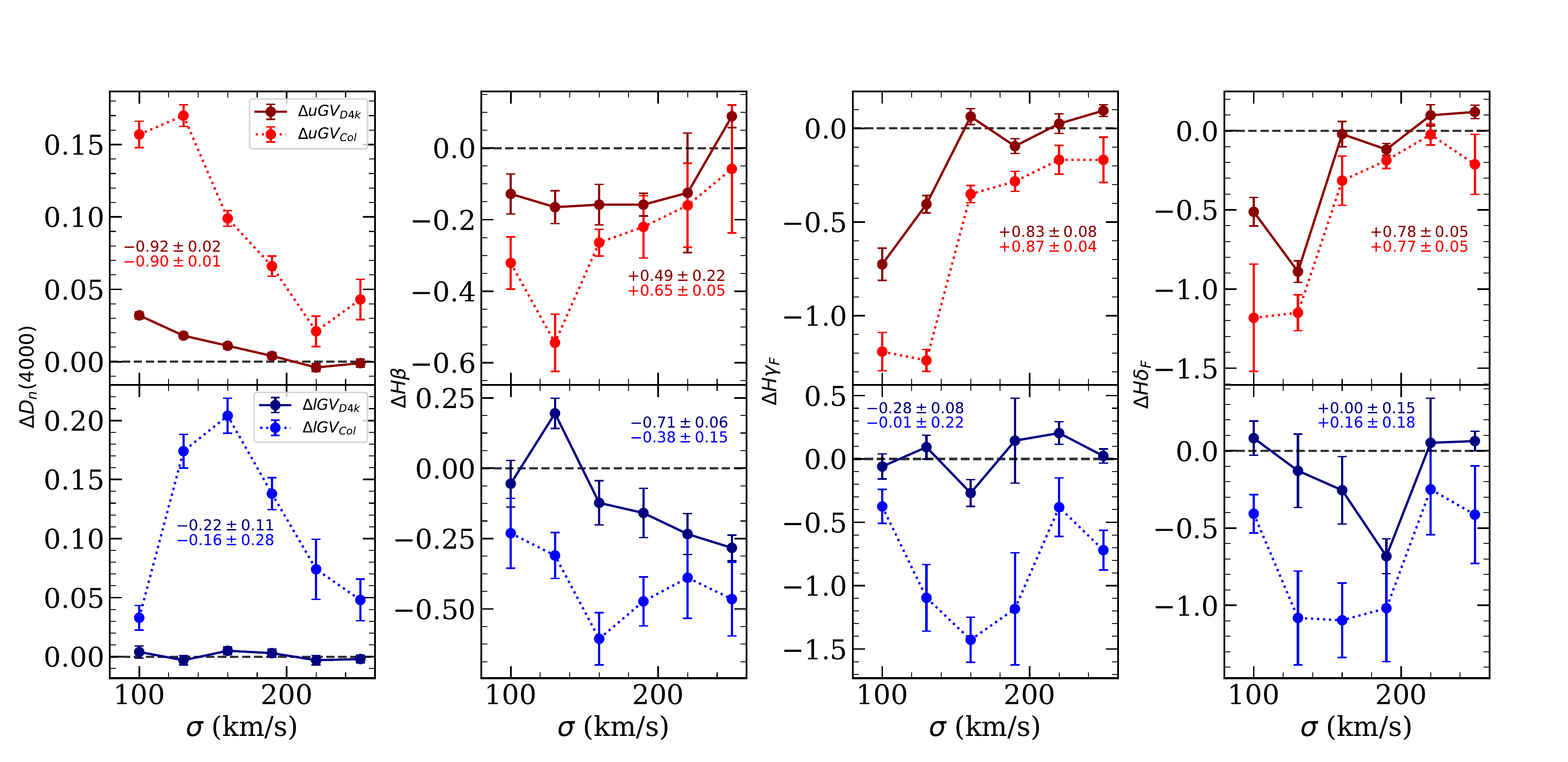}
    \caption{Differences in age-sensitive line strengths between uGV
      (top panel) and lGV (bottom panel) between Q only stacks and
      ``full'' stacks, comprising SF,
      LINERS and Q galaxies. The light red and blue dotted data shows the
      difference for colour selection, while dark red and blue solid
      lines show the difference in the D4k 
      selection. Likewise, the numbers shown in light red and light blue
      give the Pearson correlation coefficient regarding the colour selection
      data points, while the dark red and blue text shows the
      correlation coefficient calculated for the D4k-selected sample.
      The error bars are quoted at the 1\,$\sigma$ level. }
    \label{fig:IdxDiff1}
\end{figure*}

Fig.~\ref{fig:IdxDiff1} shows the difference in the age-sensitive
indices between the Q stacks (hereafter $Q_{stack}$) and the combined
Q$+$SF$+$LINER stacks (hereafter $All_{Stack}$).  The top and bottom
panels show the difference in results for uGV and lGV galaxies,
respectively.  Solid and dashed lines show the results for D4k and
colour selection, respectively.  In uGV galaxies, a decrease in the
difference of $D_n(4000)$ with respect to velocity dispersion is
evident, for both D4k (pcc=$-0.92\pm0.02$) and colour selection
(pcc=$-0.90\pm0.01$).  This trend can be attributed to the lower
contribution of star-forming galaxies as velocity dispersion increases
(see Table~\ref{tab:fractions}).  It should be noted that the
variation in $\Delta D_n(4000)$ (leftmost panels) for the D4k
selection should be, by construction, minimal, as we are constraining
the $D_n(4000)$ index within certain values. Note, in contrast, the
different trends found in the Balmer indices. For the uGV, inclusion
of SF galaxies increases Balmer absorption in both definitions of the
GV, as expected from the decrease in average age.  Furthermore, the
values of pcc show a stronger correlation/anti-correlation in uGV in
comparison to lGV; this effect can be attributed to a lower percentage
of SF and AGN galaxies in uGV in contrast to lGV with increasing
velocity dispersion.  We find a more homogeneous distribution of
stellar populations when GV galaxies are selected according to D4k,
whereas the colour-based selection produces larger differences in the
indices in both uGV and lGV.

\begin{figure*}
    \centering
    \includegraphics[width=\linewidth, height=70mm]{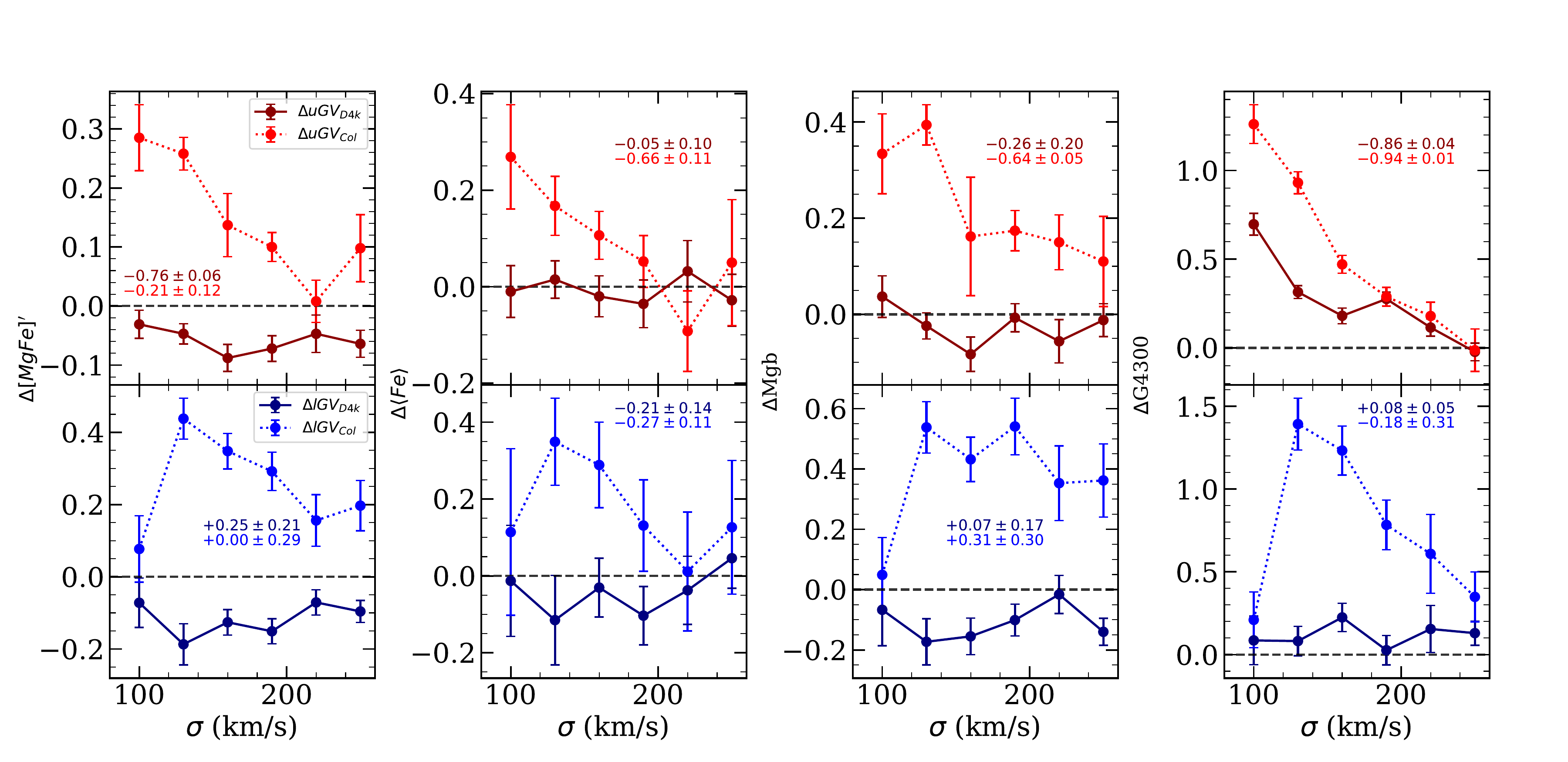}
    \caption{Same as Fig. \ref{fig:IdxDiff1} but here we show
      metallicity sensitive indices and G4300. }
    \label{fig:IdxDiff2}
\end{figure*}

Fig.~\ref{fig:IdxDiff2} shows the results for the remaining line
strengths.  Note the age-sensitive G4300 behaves similarly to
$D_n(4000)$, as in Fig.~\ref{fig:LineS}.  Both Figs.~\ref{fig:IdxDiff1} and
\ref{fig:IdxDiff2} show that in all cases, the line strength
variations are significantly smaller for the D4k selection,
confirming that a more homogeneous distribution of GV galaxies is
produced when selecting the sample on break strength. A colour-based
selection, even after a careful dust correction is applied, produces a
more complex mixture of galaxies, regarding nebular activity. We argue that the D4k selection
results in a more representative sample of transitioning galaxies from
the BC to the RS. Note that for the metallicity sensitive indices, the
difference between stacks gives a negative value, that may suggest 
lower average metallicities when stacking only quiescent galaxies. Note
that, e.g. [MgFe]$^\prime$ also increases with stellar age, so that an
age-dependent explanation would need to invoke {\sl younger}
populations in Q galaxies. In contrast, the colour selection yields
positive variations in the metal-sensitive indices, as expected if
$All_{stack}$ is dominated by younger, metal-poor, star-forming galaxies.
This result further supports the idea of population
contamination when using colour selection. This contamination seems to
affect not only the age- but also the metallicity-sensitive indices
at all values of the velocity dispersion, although the difference appears
to decrease towards higher velocity dispersion, where the contamination
from dusty BC galaxies decreases.

\section{Spectral fitting analysis}
\label{sec:CompAnalysis}

In order to look in more detail at the stellar population content of
GV galaxies, we carried out spectral fitting, by use of the {\sc
  STARLIGHT} code \citep{STARLIGHT05}, to produce best-fit {\sl
  composite} mixtures.  Variations of these mixtures across the GV
will inform us of the transition between the lGV to the uGV -- note
that depending on whether the dominant mode is quenching or
rejuvenation, it is possible to evolve {\sl in both directions}. {\sc
  STARLIGHT} performs linear superpositions of simple stellar population (SSP) spectra
supplied by the user, selecting a best fit by minimizing a
$\chi^2$ statistic with an MCMC sampler. In this paper we use
a grid of $N_*$=138 SSPs from the models of 
\citet[][hereafter BC03]{BC03}, adopting a \citet{Chabrier_2003}
IMF.  Our grid consists of 28 distinct stellar ages and 6 different
metallicities. The age ranges from 1\,Myr to 13\,Gyr, spaced
logarithmically, and the metallicity varies from [Z/H]=$-$2.3 to
$+$0.4.

We mask out the standard spectral regions where nebular emission may be
prominent. The fitting range we chose -- 3500-7500\AA -- includes a number
of age- and metallicity-sensitive regions such as those targeted in the
previous section. Spectral fitting provides an alternative approach
to individual line strength analysis; the larger amount of information,
including the stellar continuum from the NUV, optical and NIR windows, 
allows for constraints on more complex distributions of stellar ages
and metallicities.

\subsection{Luminosity-weighted parameters} \label{sec:Lum_weight_Ave}

\begin{figure*}
    \centering
    \includegraphics[width=0.45\linewidth]{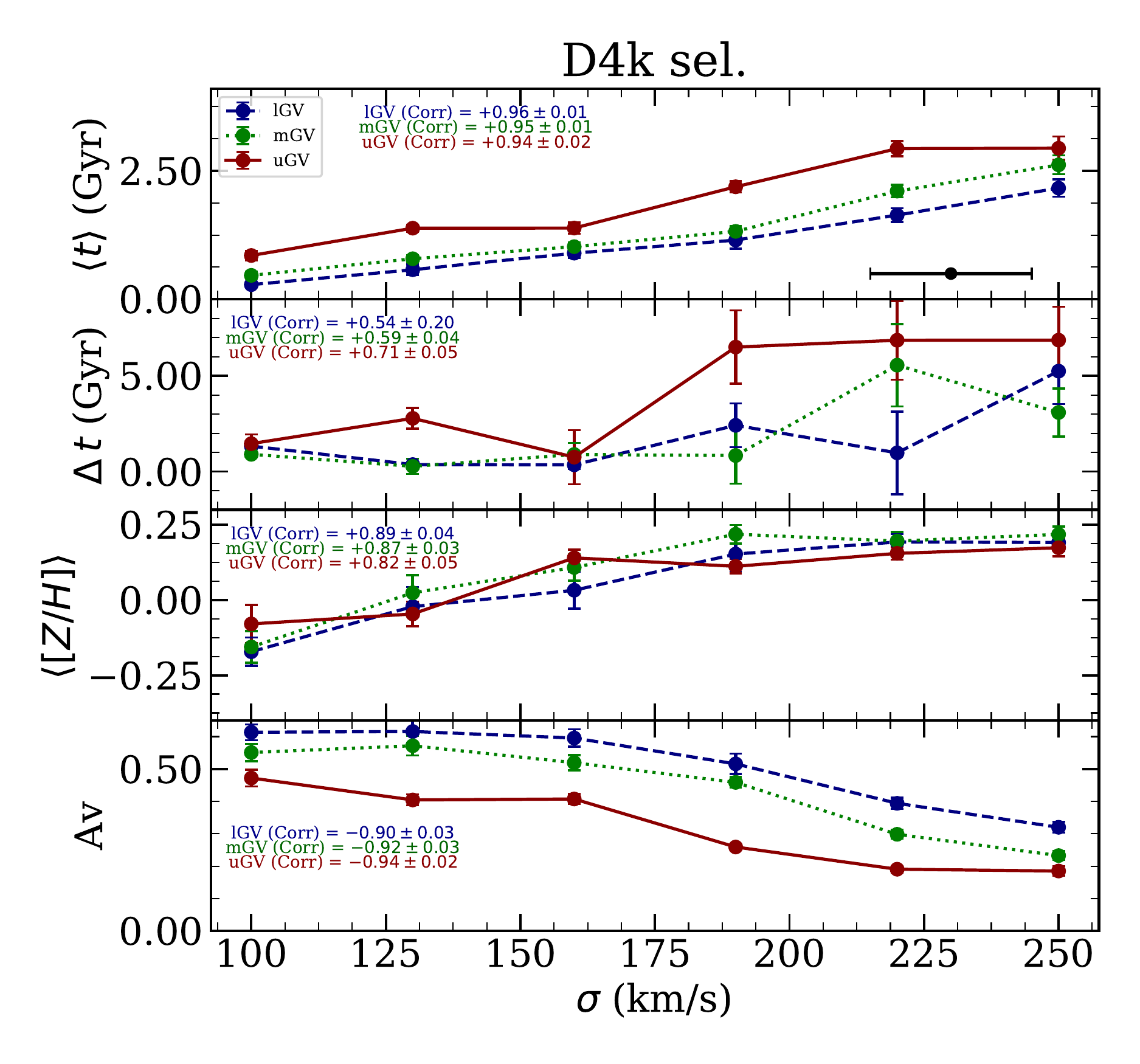}
    \includegraphics[width=0.45\linewidth]{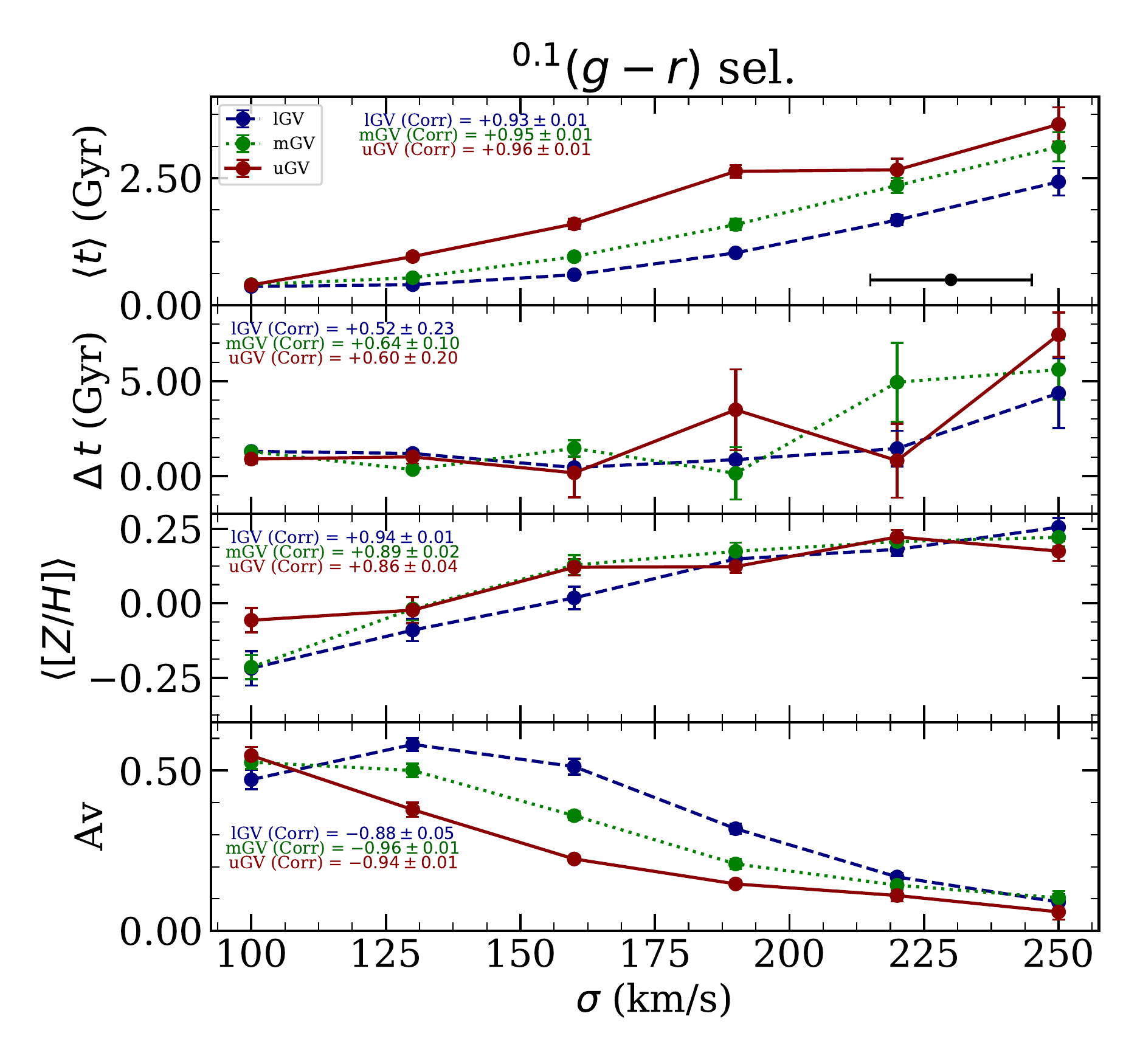}
    \caption{From top to bottom, trends with velocity dispersion of
      average stellar age, $\Delta t$ -- that acts as pseudo quenching
      timescale -- metallicity, and dust attenuation. The blue dashed,
      green dotted and red solid lines show the results for lGV, mGV
      and uGV, respectively. Results are shown for the 4000\,\AA\  break
      selection (left) and the colour selection (right) of GV galaxies. 
      The numbers in each panel quote the Pearson correlation coefficient (pcc),
      where the colours correspond to lGV, mGV and
      uGV, accordingly.}
    \label{fig:SL1LW}
\end{figure*}

Fig. \ref{fig:SL1LW} shows the luminosity-weighted averages of
some key stellar population parameters of the GV stacks
as a function of velocity dispersion. The left (right) panels show 
the D4k- (colour-) based selection, respectively.
The blue dashed, green dotted and
red solid lines represent the result for lGV, mGV and uGV, 
respectively. Each panel quotes the corresponding Pearson correlation
coefficient (pcc).
From top to bottom, we show the average stellar age, $\langle t\rangle$,  
a parameter, $\Delta t$, defined below, that keeps track of the width of the age distribution, 
the average
total metallicity, $\langle[Z/H]\rangle$, and the dust attenuation applied as a foreground
screen, $A_V$. The average age is defined as:
\begin{equation}
    \langle \log\ t\rangle = \sum_{j=1}^{N_{*}} x_j\,\log\ t_j, 
    \label{eq:tave}
\end{equation}
where $x_j$ is the normalized luminosity weight (i.e. $\sum_j x_j=1$) 
and $t_j$ is the age of the $j$-th SSP in the basis set \citep[see][for details]{STARLIGHT05}. 
The average age shows the well-established trend, where velocity
dispersion, roughly a proxy for galaxy mass, is positively correlated
with the average age \citep[e.g.,][]{AnnaGallD4k}. Galaxies on the uGV
are consistently older at all values of velocity dispersion. The age
difference between mGV and lGV is less prominent than between these
two and the uGV. A possible explanation for this trend is discussed in
Sec. \ref{sec:Dis_Phys}. The age difference between the uGV and lGV
stays in the region 0.5-1\,Gyr, a consistent result with respect to
the SSP analysis shown in Sec.~\ref{sec:SSP}.

The uncertainties of the parameter estimates are obtained by making 20
Monte Carlo realizations of each stack, by adding noise consistently
with the uncertainties in each flux bin, and re-running each one of
them through {\sc STARLIGHT}. The realizations produce a distribution
from which the standard deviation of the parameter estimates is quoted
as the uncertainty. Galaxies in the lGV have a smoother age trend 
with respect to velocity dispersion.

The second age indicator, $\Delta t$ (second panel from the top),  
is extracted from the cumulative stellar mass profile:
\begin{equation}
{\cal C}(s) = \sum_{j} x_j(t\leq s).
\end{equation}
Defining the inverse of this function as $\tau(y)\equiv{\cal C}^{-1}(y)$, we
take the time interval $\Delta t\equiv \tau(0.70) - \tau(0.30)$, i.e.
the time lapse 
spanning the epochs when the galaxy, or its progenitors, 
formed between 30\% and 70\% of its total
stellar mass. The motivation behind this definition is to provide
a GV transition interval that could be related, for instance, with
a quenching timescale. However note two important caveats. Firstly, 
if the galaxy has experienced a rejuvenation episode, the recently
formed stars will be given more weight, thus leading to high $\Delta t$
values: even if the actual quenching timescale is relatively short.
Secondly, we are dealing with stacks that include star-forming
galaxies. Therefore, our definition of $\Delta t$ should be considered
as an effective GV timescale, rather than a true quenching interval.
Previous results from the literature noted that the quenching timescale
evolves in a complex way with velocity dispersion, with an initial
increase, followed by a decrease at the massive end  \citep{QuenchMethods,Wright:19}.
Here, we find $\Delta t$ increases in all cases, in contrast with the
trends found in the line strength analysis of A19, where a
non-monotonic behaviour was found between the SSP-equivalent age difference of
galaxies on the uGV and lGV with respect to velocity dispersion.

The third panel from the top shows the average total metallicity, 
calculated in a similar manner to average age, namely:
\begin{equation}
\langle [Z/H]\rangle = \sum_{j=1}^{N_{*}} x_j [Z/H]_j,\label{eq:Zave}
\end{equation}
where $[Z/H]_j$ is the individual SSP metallicity of the $j$th
component.  A strong positive trend is also found between total
metallicity and velocity dispersion, once more in agreement with
previous studies of the general population \citep[see,
  e.g.,][]{AnnaGallD4k,Graves2010DRS}.  At the low mass end, the
colour-based selection (panels on the right), shows a consistent trend
towards a higher metallicity in the uGV with respect to the
lGV. However, this trend is less evident in the D4k selection (left
panels), which is less sensitive to contamination from dusty,
star-forming (and possibly lower metallicity) galaxies, especially in
the lGV; resulting in a weaker correlation for D4k selected samples 
in comparison to a colour-based GV. 
The D4k selection shows that, within error bars, metallicity
does not segregate within the GV at fixed velocity dispersion. This
result suggests a potential bias when selecting GV according to
colour.  The next (bottom) panels of Fig.~\ref{fig:SL1LW} show the
monotonically decreasing trend of dust attenuation with $\sigma$,
consistent with previous studies \citep[e.g.,][]{2014Barb}, featuring
a clear stratification from lGV to uGV, with mGV galaxies having, once
more, properties closer to lGV galaxies. This information can also
help assess potential biases related to the dust correction needed in
the colour-based selection. It is worth mentioning the small
difference in $A_V$ at the lowest velocity dispersion in colour-based
GV galaxies. One could argue that the dust correction could be partly
responsible for this result, whereas the D4k selection shows the $A_V$
stratification between uGV, mGV and lGV at all values of $\sigma$.

\begin{figure*}
  \centering
  \includegraphics[width=0.45\linewidth]{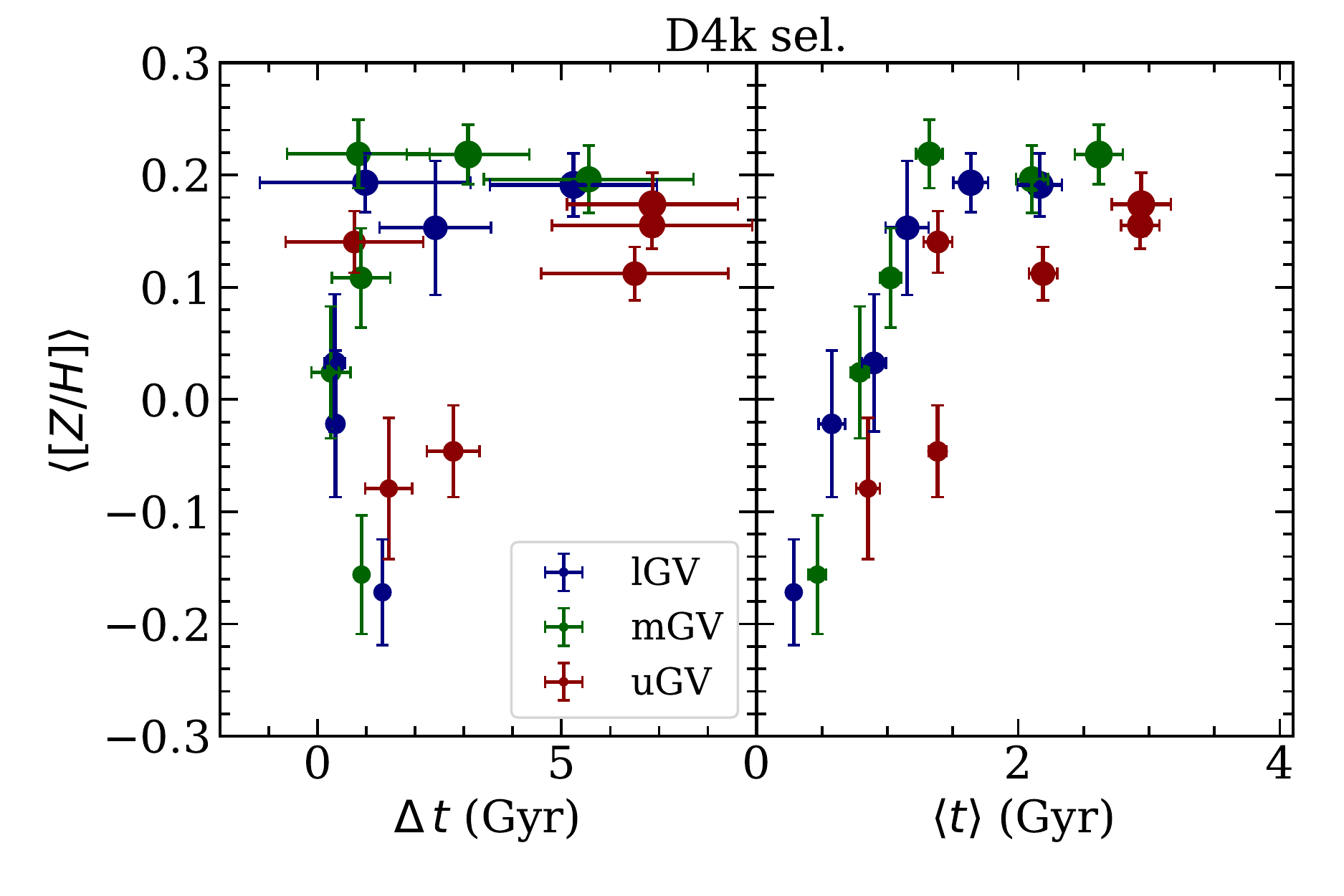}
  \includegraphics[width=0.45\linewidth]{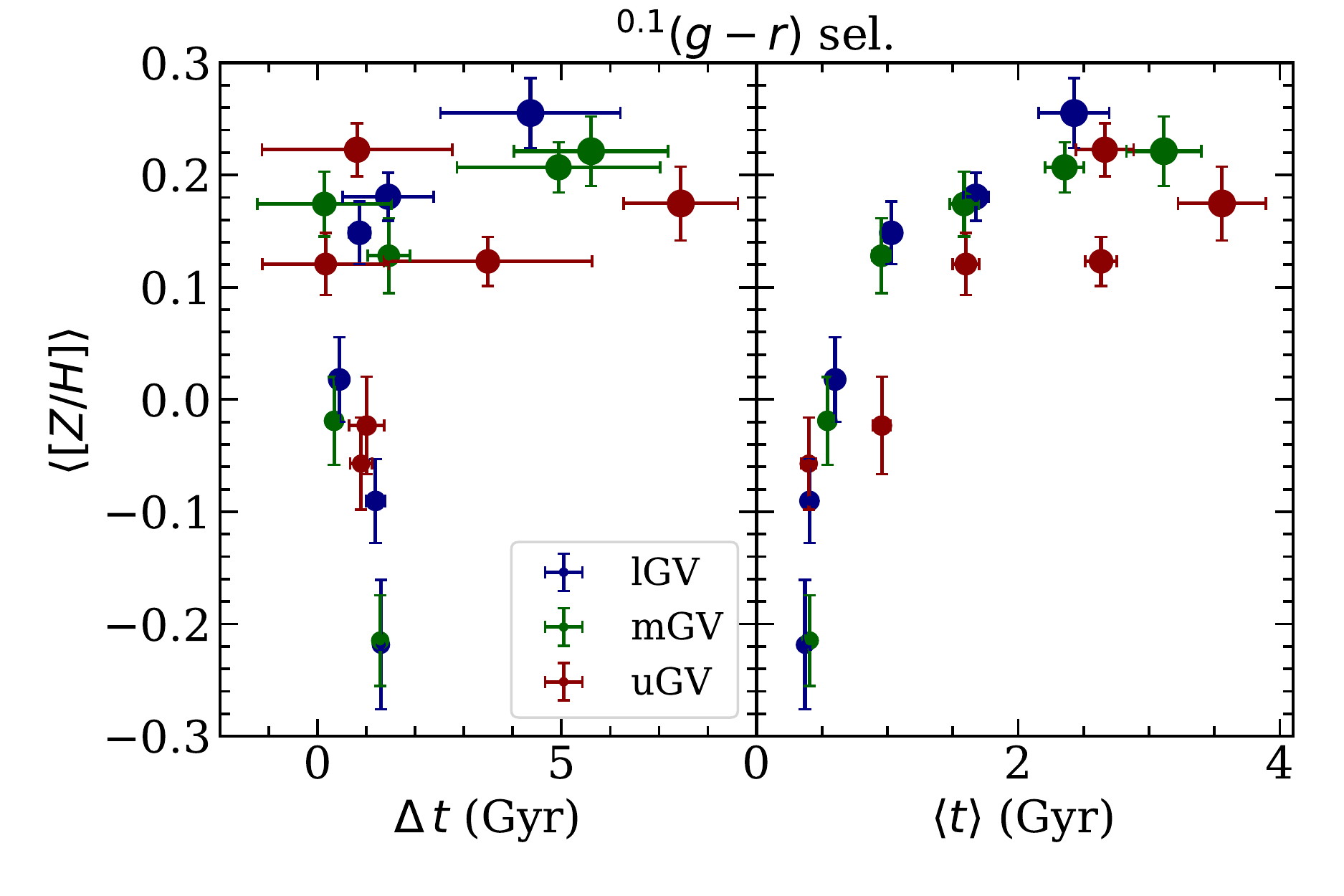}
  \caption{Trends between metallicity and either $\Delta t$, or
    average age.  The blue, green and red data points show the results
    for lGV, mGV and uGV respectively. The marker size follows
    velocity dispersion. The uncertainty in average metallicity
    remains constant, whereas the error bar in $\Delta t_Q$ increases
    with average metallicity. Results are shown for the
    4000\,\AA\ break selection (left) and the colour selection (right)
    of GV galaxies.}
  \label{fig:SL2LW}
\end{figure*}

Fig.~\ref{fig:SL2LW} shows the variation in average
metallicity as a function of $\Delta t$ and average age, as labelled.
The results of a D4k (colour) selection of GV galaxies is shown on the
left (right) panels, following the same colour coding as the previous
figure to represent the uGV, mGV, lGV stacks.
The marker size maps velocity dispersion. Focusing on the D4k selected sets,
we find two different trends in the sample:
i) At low metallicity
($\langle[Z/H]\rangle\lesssim +0.1$), GV galaxies have short
$\Delta t\,\,(\lesssim$3\,Gyr);
ii) At higher metallicity, GV galaxies have a broader distribution of
$\Delta t$, a result indicative of 
rejuvenation. In this region, galaxies have higher velocity dispersion
and feature older stellar populations. Moreover, lGV and mGV galaxies have
relatively shorter $\Delta t$, with respect to uGV systems, that have 
$\Delta t\gtrsim$5\,Gyr. Therefore, a higher fraction of uGV galaxies
at the massive end appear to have undergone more substantial episodes
of rejuvenation.  The standard age-metallicity relation can be found
in the figure, with a significant stratification towards
older populations at fixed metallicity in uGV galaxies.
This effect is more pronounced in the D4k
selection, whereas the colour-based selection produces a more complex
mixture at low velocity dispersion, as expected from the contribution
of dust, more prevalent at this end of the distribution.

\subsection{Star Formation History (SFH)} \label{sec:SFH_STARLIGHT}

\begin{figure*}
    \centering
    \includegraphics[width=145mm]{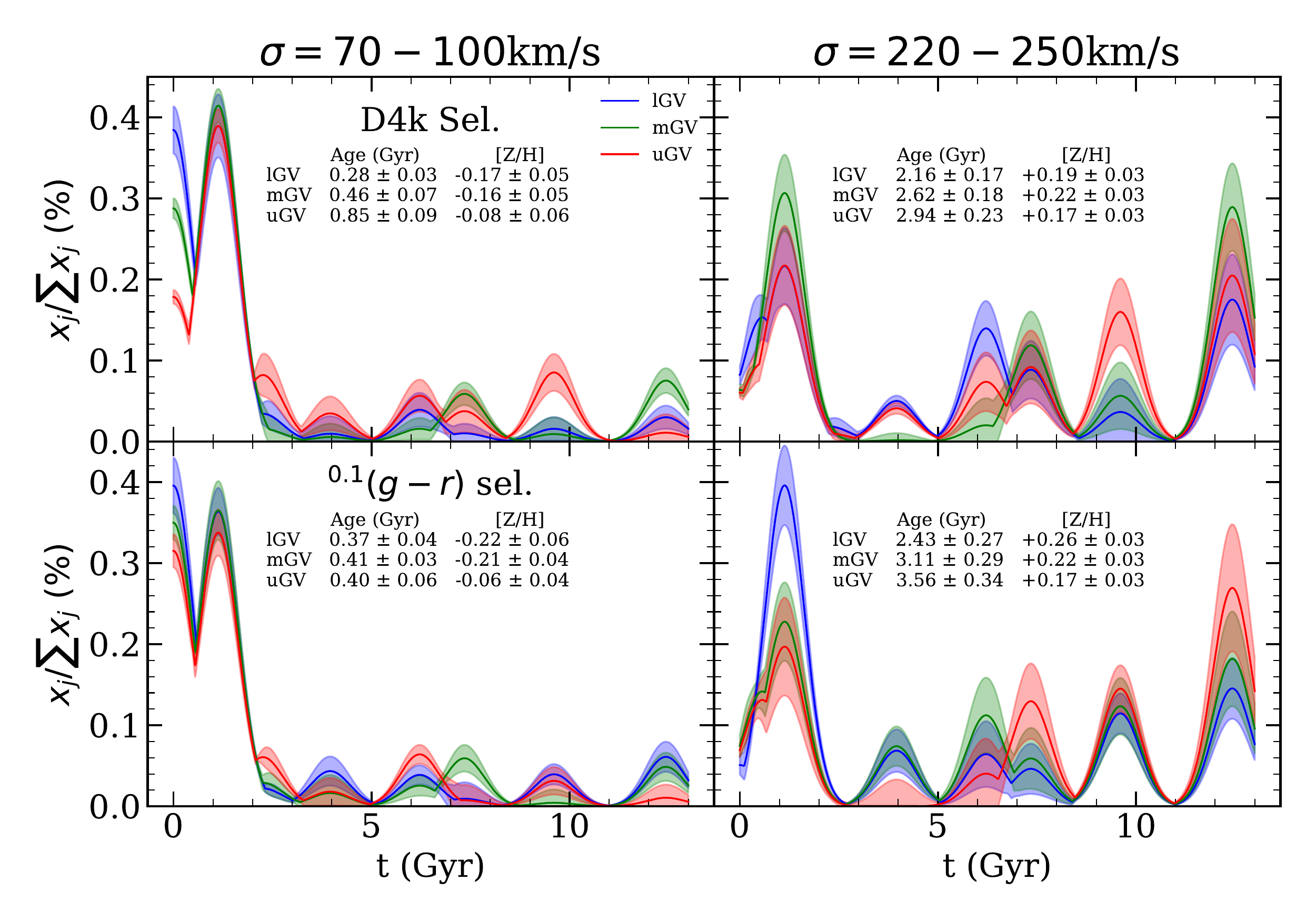}
    \caption{The star formation history is shown for both $D_n(4000)$ (top) and colour (bottom)
      selected GV galaxies. The left and right sides show the SFH in the 
      lowest velocity dispersion bin $70<\sigma<100\, $km/s and
      the highest velocity dispersion bin $220<\sigma<250\, $km/s,
      respectively. The 
      blue, green and red curves represent the SFHs of lGV, mGV and
      uGV galaxies, respectively.}
    \label{fig:LWSFH}
\end{figure*}

The next step in the analysis of the spectral fitting constraints is
the star formation history, namely the distribution of
SSP weights, ${x_j}$ as a function of age -- i.e. marginalized with
respect to metallicity.  Note that individual SFHs constrained on a
galaxy by galaxy basis are rather uncertain with any population
fitting code, and {\sc STARLIGHT} is no exception.  However as we are
dealing with stacked spectra covering a large number of galaxies, we
can assume that the derived SFHs represent statistical trends in the
various regions of the GV probed here.
Fig.~\ref{fig:LWSFH} shows the luminosity-weighted output
for lGV (blue), mGV (green) and uGV (red). The
histograms bin the age intervals in a linear manner. The
average age and metallicity, quoted in each panel, have been calculated
using the SSP ages of the basis set, along with their corresponding weights, 
following equations \ref{eq:tave} and \ref{eq:Zave}. The left and right panels show the
results at the lowest and highest velocity dispersion bins,
70$<\sigma<$100\,km/s and 220$<\sigma<$250\,km/s,
respectively. The upper (lower) panels show the results for the D4k
(colour) selection, as labelled. Note these results are robust
regarding the {\sl relative} weight contributions, whereas absolute
estimates may carry larger uncertainties.
To assess the statistical
uncertainty, we perform for each stack a Monte Carlo set of 20
realizations of spectra with the same flux distribution and noise
compatible with the stack under consideration, following an identical
methodology. The shaded regions in the figure show the expected
uncertainty from this comparison. Note that spectral
fitting inherently constrains luminosity-weighted properties. A
translation into mass-weighted values unavoidably
carries additional uncertainties, related to the mapping from luminosity
into stellar mass. For instance, a recent episode of star formation
can bias the results as the hot, massive stars present in young
populations contribute significantly more than their cool, low-mass
(although equally young) counterparts.

\begin{figure*}
  \includegraphics[width=0.45\linewidth]{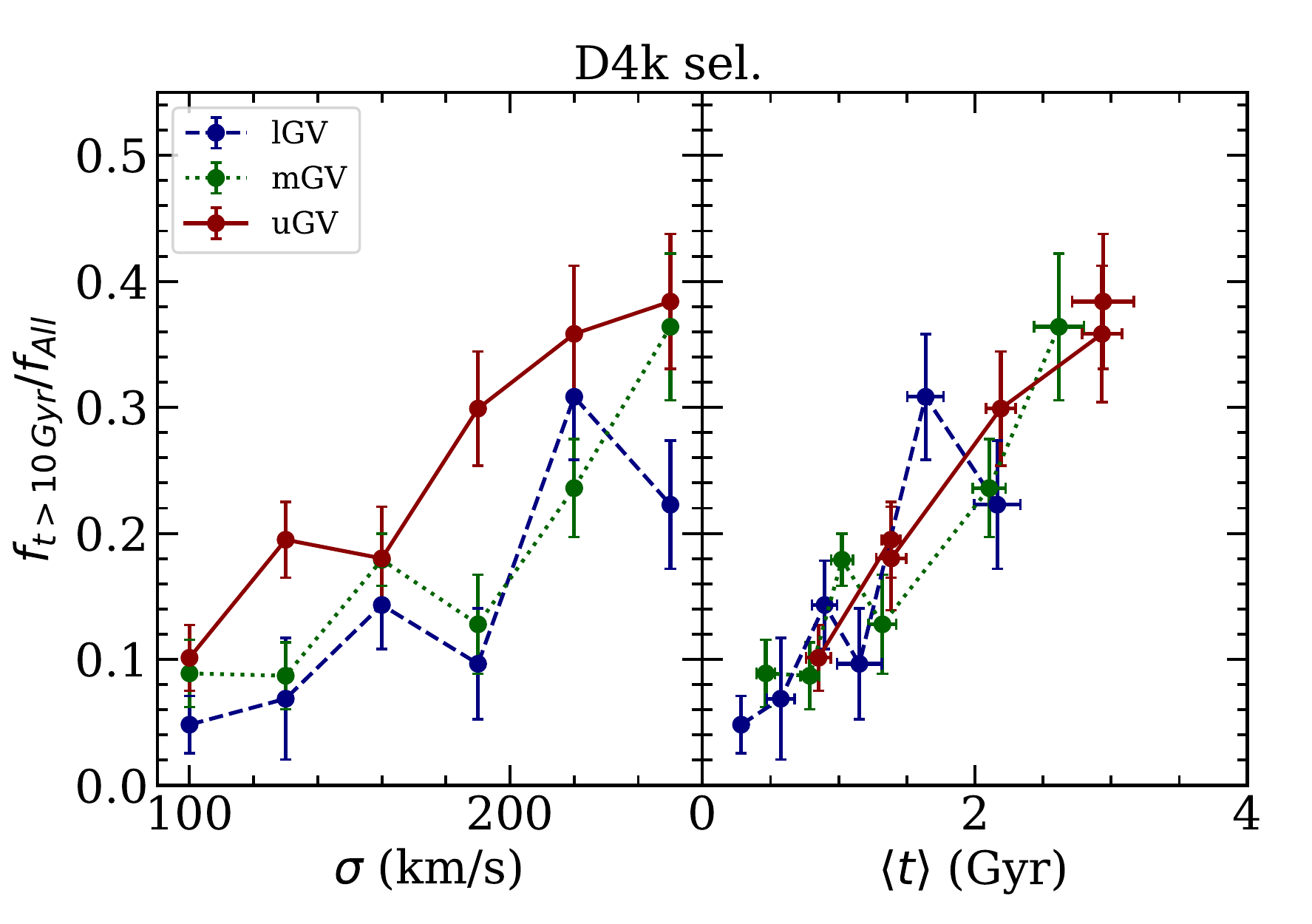}
  \includegraphics[width=0.45\linewidth]{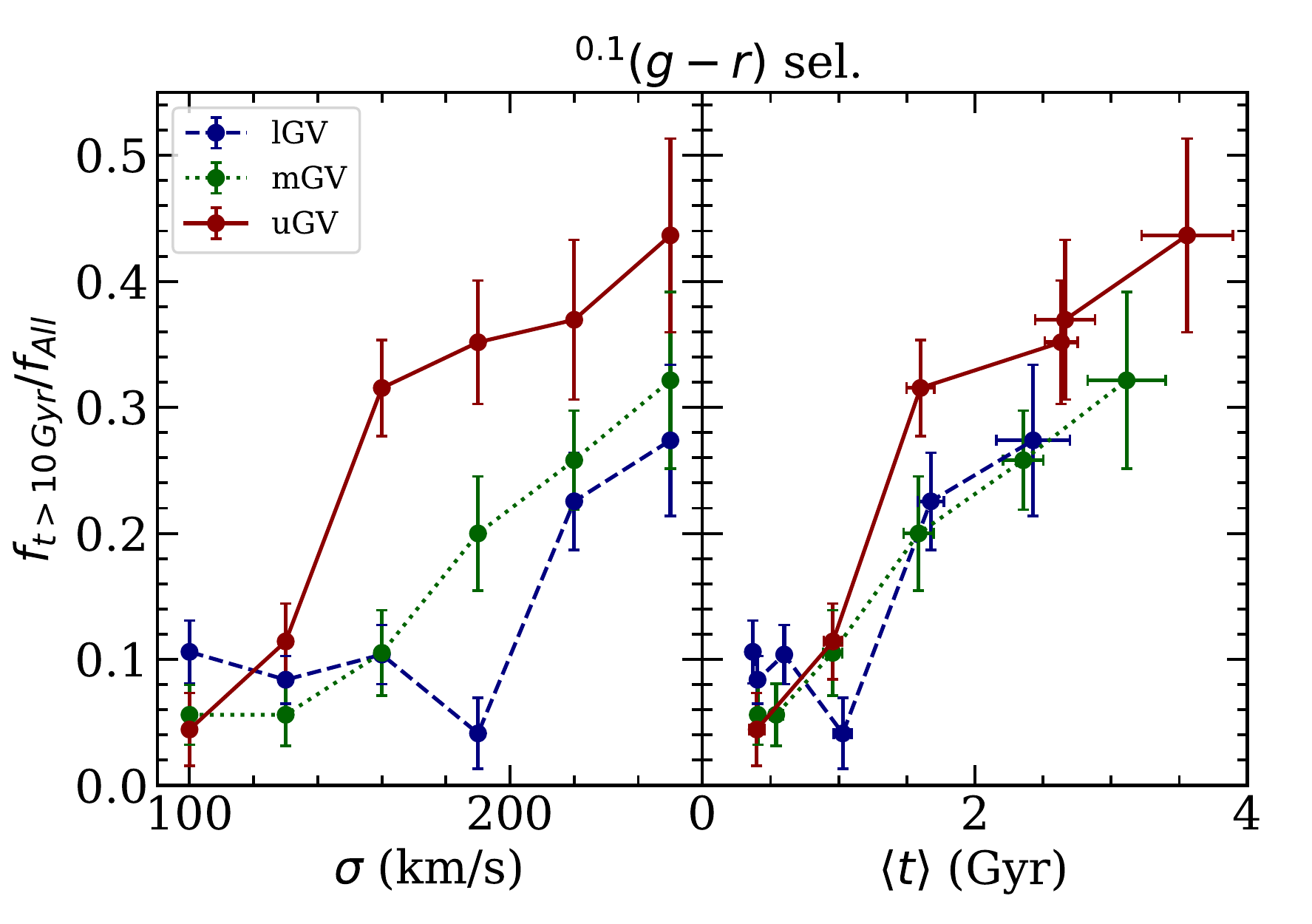}
  \caption{Relation between the mass fraction in old stellar
    populations ($\gtrsim$10,Gyr), with respect to velocity dispersion
    (left) and average age (right).  The blue, green and red lines
    correspond to lGV, mGV and uGV galaxies, respectively.  Results
    are shown for the 4000\,\AA\ break selection (left) and the colour
    selection (right) of GV galaxies.}
  \label{fig:SL3LW}
\end{figure*}

At low velocity dispersion (left panels) the stellar populations are
mostly young. As we traverse the GV from lGV to uGV, more weight is given to the older
components, increasing the average age from
$\langle t\rangle\sim$0.3\,Gyr in the lGV to $\sim$1\,Gyr in the uGV,
for the D4k selection, whereas the colour-based selection produces a more
homogeneous distribution, with undistinguishable age differences
``across the valley'', and a significantly larger scatter. In all cases,
the SFHs concentrate within the most recent $\sim$2\,Gyr, with small,
subdominant old populations that become, only slightly, more prominent in the
uGV. At high velocity dispersion (right panels), the SFHs evolve towards
older components, with average (luminosity-weighted) ages $\sim
2-3$\,Gyr.

Fig.~\ref{fig:SL3LW} shows the variation of the mass fraction in old
($>$10\,Gyr) stars with respect to velocity dispersion and average
age, for the D4k (left) and colour (right) selection.  Both cases
produce similar increasing trends of the old contribution  in
the most massive galaxies and towards higher average ages. 
The
latter statement is not trivial, as this diagnostic is sensitive to
whether the age distribution changes its width with respect to average
age. The D4k selection produces consistently higher old stellar
fractions at all values of velocity dispersion, whereas the
colour-based selection, once more, shows some mixing at the low mass
end. Regardless of the selection process, uGV galaxies display more of
a difference with respect to mGV and lGV galaxies. Note that at low
velocity dispersion, the colour selection yields a lower fraction in
old stars, specially in the uGV. This is interesting as even though
both selection methods feature not too dissimilar low fractions of Q galaxies
($7.3\pm0.4$\% for D4k, and $4.2\pm 0.4$\% for colour), we see a
greater number of SF galaxies in the colour-based selection
(namely $70.0\pm 1.3$\% in the D4k selection versus $78.4\pm2.1$\% for
the colour selection). This could be a further indication of a possible
bias due to dust attenuation that causes BC galaxies to ``creep'' into the
GV \citep{Schawinski2014}. This theory is
further supported at intermediate velocity dispersion bins,
100$<\sigma<$190\,km/s, where even though there is a lower
contribution of SF galaxies in the D4k Selection of the uGV, a higher
fraction of old stars is found in the colour-based GV stacks.

\subsection{Mass-weighted population properties}

In addition to the luminosity-weighted properties presented above,
we can extend the analysis by use of the stellar mass to light ratio
($\Upsilon_\star$) provided by the
population synthesis models for each SSP. Although this translation
carries additional uncertainty, it is a way to
assess whether the older components are more dominant than expected from
a simple fit to the observations, which are inherently biased towards
the most luminous stars. Mass-weighted parameters provide a more
physical interpretation of the SFHs. The analysis is based
on models with a fixed initial mass function, namely \citet{Chabrier_2003}.
We stress that within the velocity dispersion probed by this
sample, no significant variations from a ``standard'' IMF are expected
\citep[see, e.g.,][]{IF:13,FLB:13}.

\begin{figure*}
    \centering
    \includegraphics[width=0.45\linewidth]{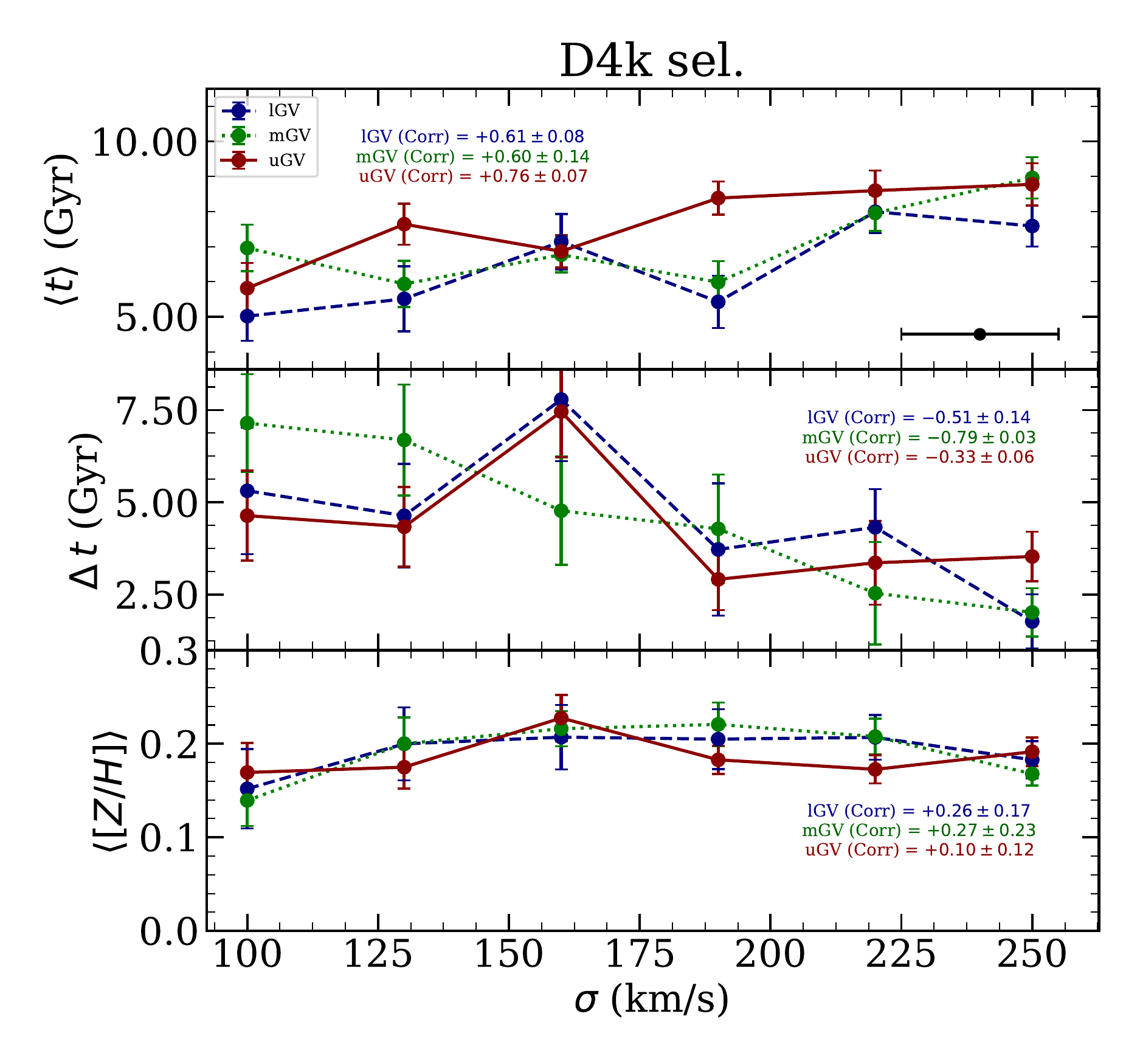}
    \includegraphics[width=0.45\linewidth]{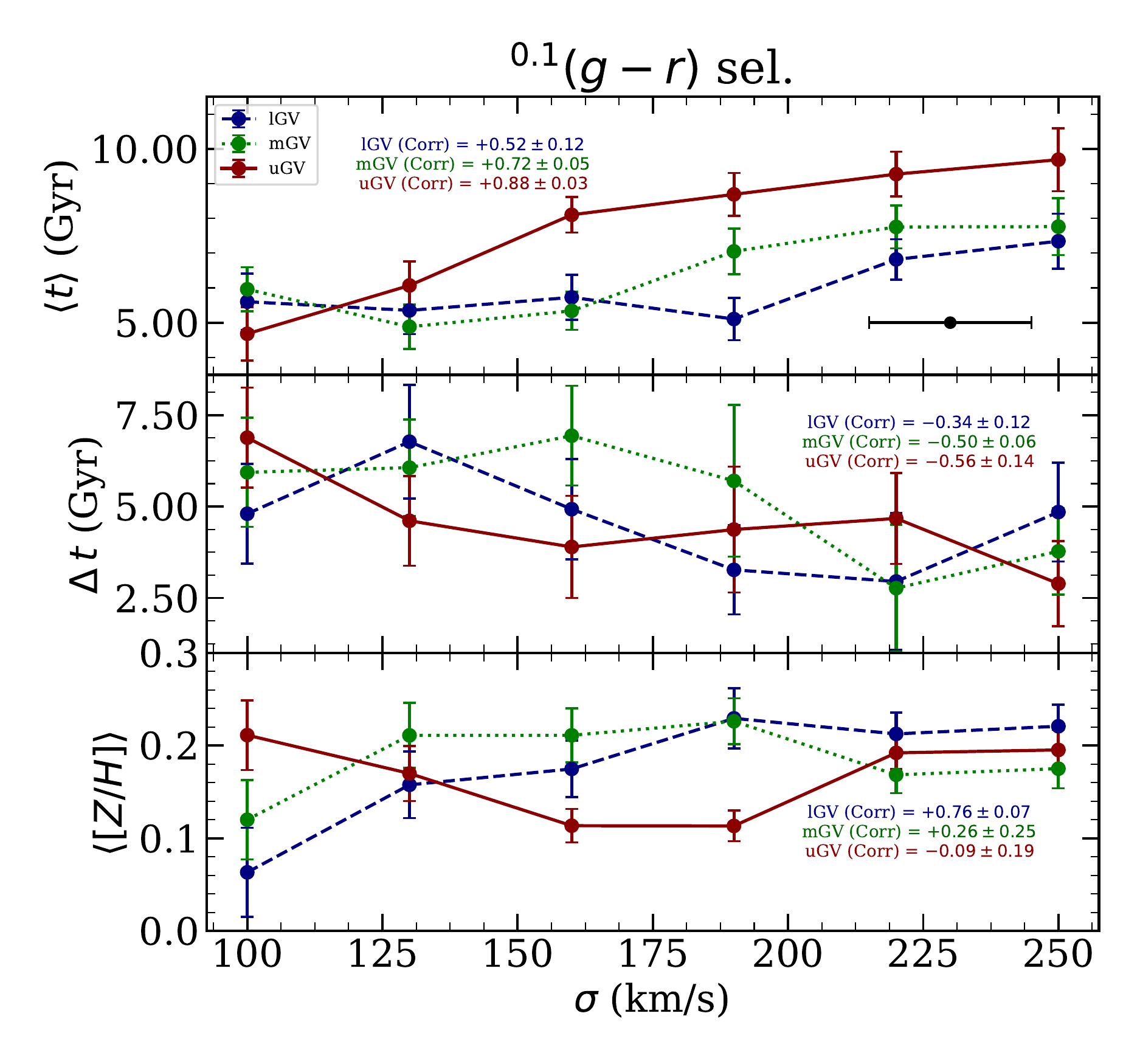}
    \caption{This is the equivalent of Fig.~\ref{fig:SL1LW}, showing
      mass-weighted parameters.}
    \label{fig:SL1MW}
\end{figure*}

Fig.~\ref{fig:SL1MW} is the equivalent of the best-fit stellar
parameters shown in Fig.~\ref{fig:SL1LW} for the mass-weighted case,
using the same line and colour coding. Similarly to the
luminosity-weighted values, a general increase is found in average age
with velocity dispersion but the absolute values are higher, as
expected, and plateau at the massive end. Regarding GV sub regions, we
also find here more affinity between lGV and mGV, whereas uGV galaxies
appear older. This result is consistent regardless of the GV selection
method, confirming that a selection based on the 4000\AA\ break
provides a homogeneous population. An increased scatter is evident in
the mass-weighted estimates, partly due to the added uncertainties
regarding the translation from light to mass.  The $\Delta t$
parameter shows a significant difference with respect to the
luminosity-weighted counterpart, with overall high values and a {\sl
  decreasing} trend with velocity dispersion, and no segregation
regarding GV location (i.e. lGV, mGV and uGV). Note, though, there is an
anti-correlation, that is weaker in comparison to
Fig. \ref{fig:SL1LW}. Estimates of average metallicity are now higher,
and appear rather flat with respect to velocity dispersion, within
error bars.

\section{Discussion} \label{sec:Discussion}

This paper looks in detail at the stellar population properties of
green valley galaxies, selected with a new methodology based on the
4000\AA\ break strength (A19), and contrasts the results with
respect to the traditional approach based on dust-corrected broadband colours. In
this section we briefly explore the implications of this analysis.

\subsection{Dust-related systematics} \label{sec:Dis_dust}

The traditional selection of GV galaxies, based on colours
derived from broadband photometry, 
is affected by dust attenuation. In order to remove
the effect of dust, a model is applied to derive a colour correction term.
However, these corrections are subject to
uncertainties that depend on  the model fitting as well as on additional
systematics related to the details of dust attenuation in galaxies. It is worth
emphasizing that the net effect of dust is to impose an effective,
wavelength-dependent attenuation law. However, in detail, it
is a result of a wide range of mechanisms involving
scattering and absorption of photons from the illuminating
source (i.e. the underlying stellar populations) by the dust
particles. The effective attenuation depends 
not only on the composition of dust, but also on
its distribution within the galaxy \citep[see, e.g.,][]{Galliano:18}.
The dust is typically concentrated around star-forming sites,
and motivates the birth cloud model, where a time-dependent
dust attenuation provides a suitable description of the net
attenuation law \citep[see, e.g.,][]{CF:00}. Observationally, star-forming galaxies 
feature a wide range of effective attenuation laws
beyond the standard ones that fit the Milky Way average
extinction curve \citep{1989Card} or the average attenuation of star-burst
galaxies \citep{Calz:00}. The variation of the parameters that describe the
attenuation law also appear to correlate \citep{Tress:18,Narayanan:18}. Therefore,
such trends can lead to significant systematics in the dust correction.

Our comparison of GV galaxies between the dust-corrected colour selection and
our proposed 4000\AA\ break strength selection enables us to assess
the role of these systematics. In appendix~\ref{app:dust} we quantify the
dependence of the two observables chosen to select GV galaxies, on
dust attenuation, using a simple attenuation law parameterised by the
colour excess.

Table~\ref{tab:fractions}, in the appendix, shows the fraction of
galaxies in the lGV and uGV, when selected according to either
D4k or colour, with the latter
shown with and without a dust correction. A graphical description of
this table can be found in fig.~3 of A19. Note the D4k-selected
GV yields a larger population of SF galaxies, than the {\sl uncorrected} colour
selection, especially towards high velocity dispersion ($\sigma\sim$200\,km/s),
along with 
a lower percentage of Q galaxies. Such a trend 
could be down to 2 reasons.  Firstly, dust could have
reddened the galaxies to different amounts, so that SF galaxies
occupy a wider region, leading to broader Gaussian PDFs when performing
a colour-based selection. Secondly, due to our
GV definition being dependent on the PDF of SF galaxies, the GV 
might have been shifted towards the RS in the colour-based approach. This explains the
large (small) percentage of Qs (SF) population in lGV $72.5\pm13.5\%$
($7.5\pm4.3\%$)and uGV $82.2\pm13.5\%$ ($8.9\pm4.4\%$), at the  highest
velocity dispersion ($220<\sigma<250\, $km/s).

The introduction of a dust correction makes the D4k- and colour-based
selections closer together (see also fig.~3 of A19), illustrating the
importance of dust correction when using colours.  In more detail,
note that the dust-corrected colour-based GV gives a lower fraction of
Q galaxies in both lGV and uGV, with respect to the D4k
selection. Although this might seem counter-intuitive, note that the
largest effect when applying the dust correction of the colours is to
increase the number of SF galaxies. Therefore, we deduce that most of
the galaxies that appear on the GV after the dust correction is
applied originate from the RS defined by the dust-uncorrected
classification -- as the effect of the correction is always to make the
colours bluer.  Another interesting trend can be seen when going from
intermediate ($\sigma$=160--190\,km/s), to high ($\sigma$=190-220\,km/s)
velocity dispersion, where we see an increase in the SF population in both D4k and
dust-corrected $^{0.1}(g-r)$ selection of lGV and uGV. Note this trend is not
seen in the dust uncorrected $^{0.1}(g-r)$ selection.

Moreover, note the difference in the line strengths between the full
GV stacks and those consisting exclusively
of Q galaxies -- shown in Figs.~\ref{fig:IdxDiff1} and \ref{fig:IdxDiff2}.
The D4k-selected sample shows a more homogeneous distribution, in
contrast with the larger variations found in the colour-based selection.
These results illustrate the highly non-trivial issue of the systematics
expected in the selection of GV galaxies, and leads us to adopt
the D4k selection as a more robust representation of the GV.

\subsection{Interpretation of the GV as a transition region}
\label{sec:Dis_Phys}

We now take the D4k-selected GV as our standard sample.  The
differences in the stellar population properties of the three
different areas of the GV, at fixed velocity dispersion, reflect the
nature of the GV as a transition phase.  The line strength results
allow us to see in a model-independent way these
variations. Fig.~\ref{fig:LineS} shows a substantial difference in the
higher order Balmer lines, especially H$\delta_F$, with respect to the
other indices. This line is especially sensitive to recent episodes of
star formation \citep[see, e.g.,][]{Martin:07} and thus may imply,
when considering stacked spectra, that the contribution from rejuvenated galaxies
dominates the flux, especially at high velocity dispersion.
Also note that at low velocity dispersion, the trend of H$\delta_F$ is smoother,
tentatively meaning that a smoother decaying (or truncated) star
formation is in operation, suggesting a quenching mechanism -- of
otherwise younger populations -- at the low-mass end.

\begin{figure}
    \centering
    \includegraphics[width=0.9\linewidth, height=80mm]{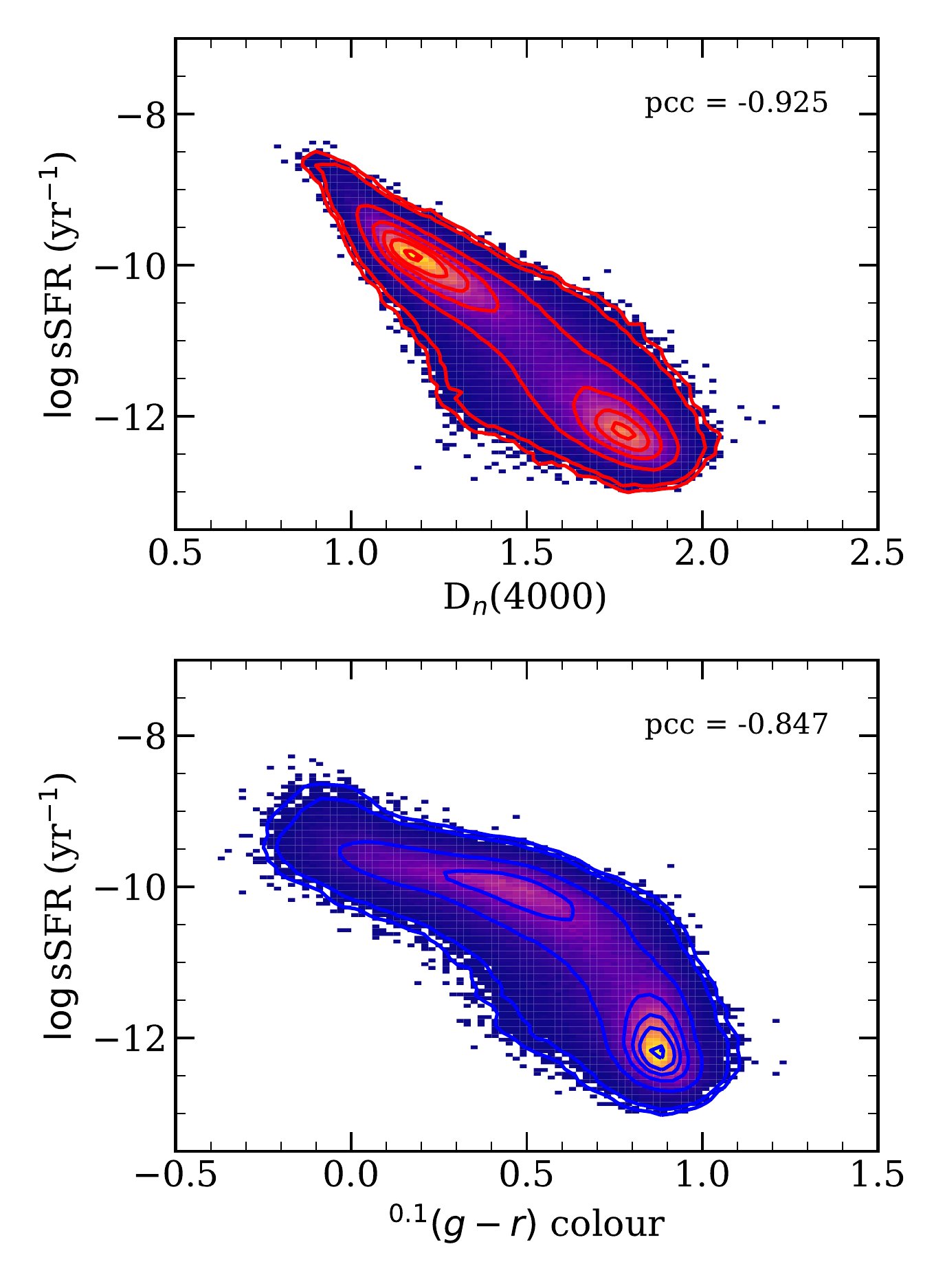}
    \caption{Relation between the sSFR and D4k $({\it top})$ and sSFR
      and dust-corrected colour $({\it bottom})$. The values denoted
      by pcc give the Pearson correlation coefficient. Note D4k shows
      a stronger (anti-)correlation with sSFR than colour.}
    \label{fig:sSFR}
\end{figure}

Further supporting the results from line strength analysis, the
population constraints based on spectral fitting
(Fig.~\ref{fig:SL1LW}), give further evidence towards rejuvenation at
high velocity dispersion: the parameter $\Delta t$ is defined as an
estimator of the width of the stellar age distribution, and stays at
$\Delta t\lesssim$2\,Gyr for $\sigma<$170\,km/s (luminosity-weighted)
-- corresponding to a more compact age distribution -- followed by
very large values of $\Delta t$ at the massive end, as expected from
the presence of two or more disjoint star formation events across
cosmic time, expected when rejuvenation produces a significant young
component.  Moreover, Fig.~\ref{fig:SL3LW} emphasizes that the
fraction of old ($>$10\,Gyr) stars in uGV galaxies is high in massive
galaxies and low at small velocity dispersion. Therefore, in this
SDSS-based, low-redshift sample, quenching appears at late cosmic
times in low-mass galaxies, whereas at the massive end one can only
measure recent rejuvenation events on an otherwise old population. The
mass-weighted results (Fig.~\ref{fig:SL1MW}) feature more scatter --
as expected from the added uncertainties in $\Upsilon_\star$,when
converting the SSP contributions into mass fractions -- but
interestingly produce a {\sl decreasing} trend of $\Delta t$ with
velocity dispersion, implying that the rejuvenation events at the
massive end cannot involve a large mass fraction of young stars.
\cite{Nel:18} explored the fractional contribution of rejuvenated
galaxies in the IllustrisTNG simulation, finding that $\sim 10\%$ of
the subset of massive galaxies (M$_\star/\mathrm{M_\odot}>10^{11} \mathrm{M_\odot}$)
have undergone rejuvenation once and $\sim 1\%$ have experienced
more than one rejuvenation event. Note our analysis shows rejuvenation
to be dominant in stacked spectra at the massive end. However, the 
percentage of different types of galaxies in the stacks
(Tab.~\ref{tab:fractions}) are roughly in agreement with the numerical 
simulations. Focusing on the uGV at high velocity dispersion 
(220$<$$\sigma$$<$250\,km/s) we find 19\% are star-forming 
galaxies. If we assume that a fraction of these SF galaxies are simply
transitioning from BC to RS, with the remainder representing rejuvenation
events, i.e. galaxies that have dropped down from RS, we might find similar
fractions of rejuvenated systems as those found in the literature
\citep{Thomas:10, Nel:18}. A similar argument can be made for galaxies
in lGV and mGV. However, the fractional contribution from star-forming
galaxies is  higher, and we expect a smaller contribution from
rejuvenation events.

Regarding metallicity, we recover the standard mass-metallicity relation,
as shown in Fig.~\ref{fig:SL1LW}. Although on a speculative tone, note
the metallicity trend between uGV and lGV gets inverted between low- and
high velocity dispersion (in the luminosity weighted version). Such a
trend would be a consequence of fresh, lower metallicity gas contributing
to the rejuvenation events at the massive end of the sample.

Concerning the transition from BC to RS, previous work from the
literature indicates a rapid evolution through the GV due to its
sparsity \citep[see, e.g.,][]{2004Bald, GAMA2015}. However, this
transition time depends on morphology. For instance,
\citet{Schawinski2014} distinguish at least two
morphologically-related transition paths. Early-type galaxies are
thought to traverse the GV in a rapid manner, quenching their star
formation very quickly and moving onto the RS, while, in contrast,
late-type galaxies are expected to undergo a slower quenching
process. This is supported by \citet{2018GAMAGV2}, where they look at
structural variations with respect to colour and morphology, and argue
in favour of inside-out formation, which is related to slow quenching,
instead of a violent transformative event. Additionally, the
observational constraints from the stacked spectra give timescales
between 2 to 4\,Gyr \citep{Phillipps:19}. Using SSP-equivalent ages,
the derived transition times are similar when stacking spectra
regardless of nebular activity (Fig.~\ref{fig:SSPAge}). However, the
difference between uGV and lGV luminosity-weighted average ages, from
spectral fitting, results in a lower transition time $\sim$1.0\,Gyr.
Interpreting $\Delta t$ as a transition timescale leads to a higher
value $\gtrsim$5\,Gyr, with respect to \citet{Phillipps:19}.  Note
this is to be expected as (i) \verb|STARLIGHT| is very robust at
tracing average parameters but produce weaker constraints on the
details of the SFH; (ii) The parameter $\Delta t$ traces the transition time
in a slightly different manner, as it is very sensitive to recent
bursting episodes (see Sec. \ref{sec:Lum_weight_Ave} for details),
while the methodology adopted by \cite{Phillipps:19} uses a fixed
exponentially decaying SFH, thus making it less sensitive to
rejuvenation effects.

In addition, state of the art simulations give further support to a 
rapid transition through the GV; 
\citet{Wright:19} state that low velocity dispersion galaxies
feature relatively long quenching timescales,
$\tau_Q\gtrsim$3\,Gyr. This timescale increases with velocity
dispersion; but at the highest values of velocity dispersion, they
find a drop to $\tau_Q \lesssim$2\,Gyr.
As for the physical mechanism that produces this transition, it is
stated that in low-mass galaxies it is mostly due to processes such as
ram pressure stripping, while in more massive galaxies quenching
operates though events such as stellar feedback
\citep{Wright:19}. Finally, in the most massive systems, major
mergers and quasar-mode AGN are thought to quench star formation.

Finally note the similar behaviour between lGV and mGV, in contrast
with uGV for the D4k-selected sample. Interestingly, this behaviour
between the different GV regions was also seen by \citet{Phillipps:19}
in their study of ``green'' galaxies using sSFR. They noted a lower
transition time going from their ``lGV'' to ``mGV'' -- 
selected by sSFR$_{lGV}$/sSFR$_{mGV}\sim$1.6, giving  2\,Gyr --
with respect to their ``mGV'' to ``uGV'' -- selected
by sSFR$_{mGV}$/sSFR$_{uGV}\sim$ 2.5, giving 3--4\,Gyr.
Fig. \ref{fig:sSFR} shows the relation between D4k and sSFR $({\it top})$
and between colour and sSFR $({\it bottom})$. The relation with D4k
presents a stronger correlation (pcc=$-$0.925) compared to colour (pcc=$-$0.847), 
therefore it is not surprising that sSFR behaves in a similar
manner to the 4000\AA\ break strength. This similarity reinforces the
trends found here.

\section{Conclusions} \label{sec:Conc}

We present in this paper a detailed analysis of the recently proposed
re-definition of green valley galaxies based on the
4000\AA\ break strength (A19), along with a comparison with
the standard selection based on dust-corrected colours from broadband
photometry. We make use of a large sample of high quality spectroscopic
data from the Sloan Digital Sky Survey (SDSS). 
The new definition adopts 
the well-known spectral index D$_n$(4000) of 
\citet{Balogh:99}, whereas the colour-based approach uses 
the SDSS-defined, dust corrected colour $^{0.1}(g-r)$, i.e.
K-corrected to redshift z=0.1. The ``population'' indicator, i.e.
either D$_n$(4000) or $^{0.1}(g-r)$, is plotted against velocity dispersion ($\sigma$),
and the sample -- defined between 70 and 250\,km/s -- is split into
six bins in $\sigma$. A probability-based approach is followed, where
the star-forming and quiescent samples define a Blue Cloud (BC) and a
Red Sequence (RS), respectively, and an intermediate population,
i.e., the green valley (GV) is introduced, and further split into three
regions, lower (lGV), middle (mGV) and upper (uGV), based on the
value of the population indicator.

Our results show overall consistent properties between the new
definition, that is more resilient to potential systematics from the
dust properties, and the dust corrected selection based on colour,
with similar fractions between star-forming (SF), quiescent (Q) and
AGN galaxies, with respect to velocity dispersion (see
Table~\ref{tab:fractions} and fig.~3 in A19).  However, when studying
the stellar population property in more detail, differences are found
between these two selection criteria that may affect the
interpretation of galaxy evolution between the BC and RS regions.  For
instance, Figs.~\ref{fig:IdxDiff1} and \ref{fig:IdxDiff2} show the
difference between the line strength indices in spectral stacks
comprising either the full set of GV galaxies, or only the quiescent
subsample. The colour-based selection produces a different, more
complex population mixture, part of which may be a systematic bias
caused by dust attenuation and the subsequent correction. We also find
that the population properties of lGV and mGV galaxies are closer
together, with the uGV sample representing an altogether different
set, with older, less dusty and more extended population mixtures (see
Figs.~\ref{fig:SL1LW} and \ref{fig:SL2LW}).

The population analysis, based on spectral fitting of high quality
stacked data, reveals the standard age- and metallicity- positive
correlations with velocity dispersion (or alternatively
mass). Moreover, we define a parameter, $\Delta t$, that describes the
width of the age distribution, and find an interesting different
between low- and high-$\sigma$ galaxies (Fig.~\ref{fig:SL2LW}). The
former have rather narrow widths ($\Delta t\lesssim$0.3\,Gyr), whereas
the latter feature more extended distributions ($\Delta
t\gtrsim$4\,Gyr).  This result is consistent with the idea that at the
low-mass end, the data reveal quenching of star formation, whereas
massive galaxies display significant (luminosity weighted) late
events, which would imply rejuvenation.  The mass-weighted equivalent
(Fig.~\ref{fig:SL1MW}), although presenting more uncertainty from the
conversion of light into mass, confirms that these events do not
contribute very large amounts in terms of the mass fraction.

\section*{Acknowledgements}

Funding for SDSS-III has been provided by the Alfred P. Sloan
Foundation, the Participating Institutions, the National Science
Foundation, and the U.S. Department of Energy Office of Science. The
SDSS-III web site is {\tt http://www.sdss3.org/}.


\bibliographystyle{mnras}
\bibliography{GViD} 



\appendix

\section{Fractions according to nebular activity}

Table~\ref{tab:fractions} shows the distribution of galaxy spectra in the
upper (uGV) and lower (lGV) green valley, with respect to velocity dispersion,
split according to nebular activity, following the BPT standard
classification \citep{BPT1981}. For reference, we show three selection criteria,
from top to bottom, the 4000\AA\ break strength (adopted in this paper), the
uncorrected ($g-r$) colour evaluated at a fiducial redshift of z=0.1, and the
dust-corrected colour, all measured within the SDSS spectroscopic fibre.
A graphical version of this table can be found in
fig.~3 of A19.

\begin{table*}
  \centering
  \caption{Number of SDSS galaxy spectra in the upper and lower
    sections of the green valley, following the definition of GV based
    on the 4,000\AA\ break strength. The table also shows the
    fractional contribution (as percentages) with respect to the
    ``spectral activity'', classified as star-forming (labelled SF;
    BPT flag 1 or 2); quiescent (labelled Q; BPT flag $-1$), or Active
    Galactic Nucleus, including LINER emission (labelled AGN; BPT flag 4 or 5)}
  \label{tab:fractions}
    \begin{tabular}{r|rccc|rccc} 
      \hline
      $\sigma$ (km/s) & No. Gal & f(SF)$\%$ & f(Q)$\%$ & f(AGN)$\%$ & No. Gal & f(SF)$\%$ & f(Q)$\%$ & f(AGN)$\%$ \\
      \hline
      \multicolumn {1}{c}{} & \multicolumn{4}{|c|}{lower GV} & \multicolumn{4}{c}{upper GV}  \\
      \hline
      \multicolumn {9}{c}{D4k selection} \\
      \hline
  70--100 & 4265 & 95.1$\pm$1.5 &  1.3$\pm$0.2 &  3.6$\pm$0.3 & 3845 & 70.0$\pm$1.3 &  7.3$\pm$0.4 & 22.6$\pm$0.8\\
 100--130 & 2369 & 77.8$\pm$1.8 &  3.2$\pm$0.4 & 19.1$\pm$0.9 & 2627 & 42.2$\pm$1.3 & 17.3$\pm$0.8 & 40.5$\pm$1.2\\
 130--160 & 1078 & 45.8$\pm$2.1 &  8.2$\pm$0.9 & 46.0$\pm$2.1 & 1344 & 24.5$\pm$1.3 & 25.7$\pm$1.4 & 49.8$\pm$1.9\\
 160--190 &  500 & 26.6$\pm$2.3 & 17.0$\pm$1.8 & 56.4$\pm$3.4 &  631 & 13.8$\pm$1.5 & 39.5$\pm$2.5 & 46.8$\pm$2.7\\
 190--220 &  190 & 33.2$\pm$4.2 & 25.3$\pm$3.6 & 41.6$\pm$4.7 &  254 & 14.6$\pm$2.4 & 51.2$\pm$4.5 & 34.3$\pm$3.7\\
 220--250 &  118 & 24.6$\pm$4.6 & 28.8$\pm$4.9 & 46.6$\pm$6.3 &  117 & 19.7$\pm$4.1 & 47.0$\pm$6.3 & 33.3$\pm$5.3\\
      \hline
      \multicolumn {9}{c}{colour selection $^{0.1}(g-r)$ (no dust correction)} \\
      \hline
  70--100 & 3384 & 95.7$\pm$1.7 &  1.7$\pm$ 0.2 &  2.6$\pm$0.3 & 3174 & 83.9$\pm$1.6 &  7.7$\pm$ 0.5 &  8.3$\pm$0.5\\
 100--130 & 1844 & 81.1$\pm$2.1 &  8.5$\pm$ 0.7 & 10.5$\pm$0.8 & 1824 & 56.2$\pm$1.8 & 25.4$\pm$ 1.2 & 18.4$\pm$1.0\\
 130--160 &  813 & 52.6$\pm$2.5 & 24.8$\pm$ 1.7 & 22.5$\pm$1.7 &  953 & 25.6$\pm$1.6 & 52.4$\pm$ 2.3 & 22.0$\pm$1.5\\
 160--190 &  311 & 30.5$\pm$3.1 & 44.7$\pm$ 3.8 & 24.8$\pm$2.8 &  353 & 16.7$\pm$2.2 & 65.4$\pm$ 4.3 & 17.8$\pm$2.2\\
 190--220 &  103 & 18.4$\pm$4.2 & 61.2$\pm$ 7.7 & 20.4$\pm$4.4 &  110 & 10.0$\pm$3.0 & 74.5$\pm$ 8.2 & 15.5$\pm$3.7\\
 220--250 &   40 &  7.5$\pm$4.3 & 72.5$\pm$13.5 & 20.0$\pm$7.1 &   45 &  8.9$\pm$4.4 & 82.2$\pm$13.5 &  8.9$\pm$4.4\\
      \hline
      \multicolumn {9}{c}{colour selection $^{0.1}(g-r)$ (dust corrected)} \\
      \hline
  70--100 & 3534 & 94.5$\pm$2.3 &  0.8$\pm$0.2 &  4.7$\pm$0.3 & 3233 & 78.4$\pm$2.1 &  4.2$\pm$0.4 & 17.6$\pm$0.8\\
 100--130 & 1623 & 69.2$\pm$2.7 &  3.8$\pm$0.5 & 27.0$\pm$1.5 & 1722 & 44.8$\pm$1.9 & 13.6$\pm$0.9 & 41.6$\pm$1.8\\
 130--160 & 1028 & 40.8$\pm$2.4 &  6.7$\pm$0.8 & 52.5$\pm$2.8 & 1194 & 22.2$\pm$1.5 & 22.7$\pm$1.5 & 55.1$\pm$2.7\\
 160--190 &  487 & 23.8$\pm$2.5 & 12.3$\pm$1.7 & 63.9$\pm$4.6 &  586 & 14.0$\pm$1.6 & 32.6$\pm$2.7 & 53.4$\pm$3.7\\
 190--220 &  231 & 31.6$\pm$4.2 & 10.0$\pm$2.2 & 58.4$\pm$6.3 &  266 & 24.1$\pm$3.3 & 42.1$\pm$4.7 & 33.8$\pm$4.1\\
 220--250 &  110 & 27.7$\pm$5.6 & 20.9$\pm$4.8 & 51.8$\pm$8.5 &  125 & 25.6$\pm$5.1 & 36.8$\pm$6.3 & 37.6$\pm$6.4\\
      \hline
    \end{tabular}
\end{table*}

\section{The effect of dust}
\label{app:dust}

\begin{figure}
    \centering
    \includegraphics[width=0.9\linewidth]{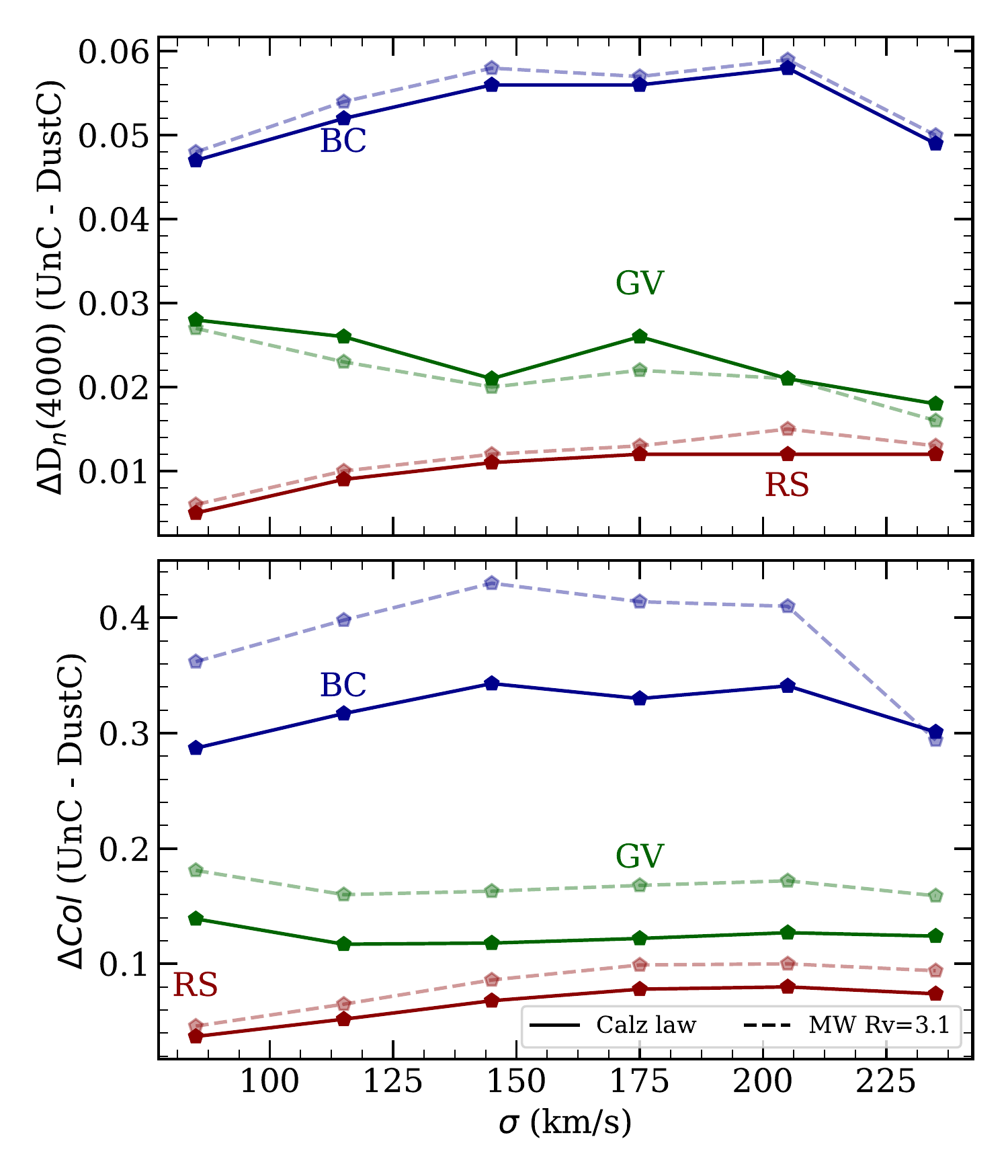}
    \caption{Dependence of 4000\AA\ break strength (top) and ($g-r$) colour (bottom)
      on dust attenuation. The blue, green and red data points show the difference
      in 4000\AA\ break strength (top) and colour (bottom), after correction
      for dust attenuation. The solid and
      dashed lines represent the two attenuation laws considered, 
      \citet{Calz:00} and \citet{1989Card}, respectively.
    }
    \label{fig:DustCorr}
\end{figure}

To illustrate the dependence of the 4000\AA\ break and
colour-selection of GV galaxies with respect to dust attenuation, we compare in
Fig.~\ref{fig:DustCorr} the difference between the trends found for
the location of the BC (blue); GV (green) and RS (red), as a function
of velocity dispersion. Each one corresponds to the mean of the PDF
associated to each sub-population (see Sec.~\ref{ssec:GVDef}). The top and
bottom panels display the difference between BC, GV and RS with and without
dust correction for D$_n$(4000) and colour, respectively. To test for
systematics we apply either a Milky Way extinction law
\citep[][dashed lines]{1989Card} or a \citet[][solid lines]{Calz:00} law. 
Note the small difference when using the
selection based on break strength (top panels), especially on GV
galaxies, at the level $\Delta D_n(4000)<$0.03, whereas the colours
(bottom panels) are not only substantially affected -- with correction
terms comparable to the actual separation between BC and RS, but are
also heavily dependent on the attenuation law adopted,
and thus prone to systematics from the variance regarding
the details of dust composition and geometry.

\section{The effect of aperture} \label{app:aperture}

\begin{figure}
    \centering
    \includegraphics[width=0.9\linewidth]{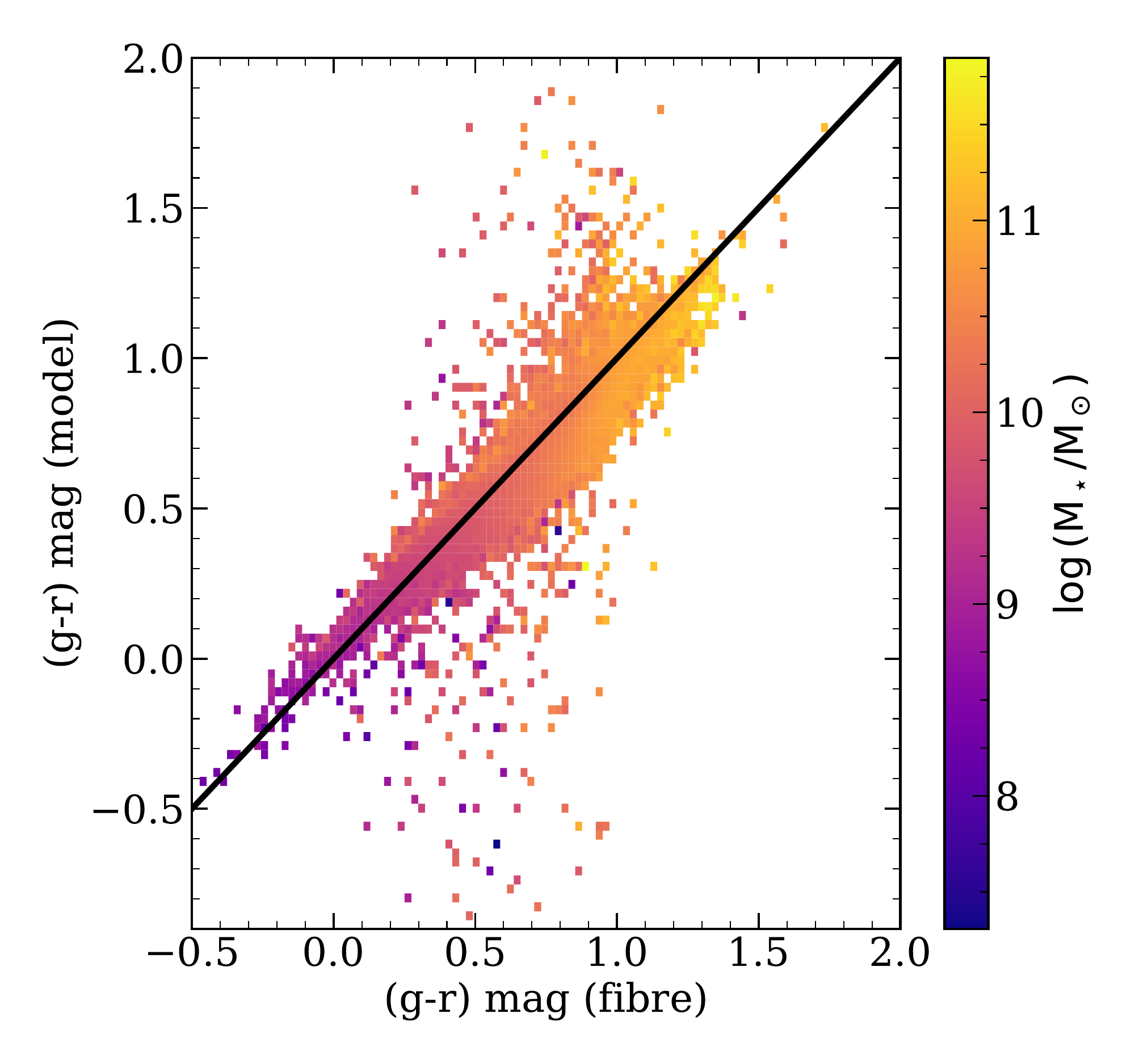}
    \caption{Relation between the colour measured inside the 3\,arcsec fibre,
      and the one from a surface brightness modelling to the whole galaxy
      (i.e. determined from the official SDSS model magnitudes).
      The data are colour coded with respect to stellar mass (colour bar on
      the right). The solid black line illustrates a 1:1 correspondence.}
    \label{fig:Ap_Ef}
\end{figure}

This study focuses on the use of spectroscopic data of the classic
SDSS dataset, therefore confined to a 3\,arcsec diameter fibre. For
consistency, our comparison with respect to colour, $^{0.1}(g-r)$,
is done within the same aperture. This could
introduce a bias with respect to stellar mass (or velocity dispersion),
as more massive, and generally larger, galaxies would have a lower fraction of
the light from their stellar populations inside the fibre.
To assess this bias, we use the JHU/MPA catalogue \citep{2003KaSM}
and compare the colour within the aperture with the modelled colour for the
whole galaxy (Fig.~\ref{fig:Ap_Ef}). The difference between these two 
definitions of colour shows a  minimal trend with stellar mass.  Due to D$_n$(4000)
tracing the average age in a similar manner to colour, we expect a similar,
minor, difference between D$_n$(4000) inside the fibre and D$_n$(4000)
measured in the whole galaxy.

Though there is no bias in relation to stellar mass, we do see an
offset between the colour in the fibre and that of the whole
galaxy. Therefore, a similar behaviour in D$_n$(4000) is expected. The
inconsistency in colour between fibre and whole galaxy is partially
due to biases from surface brightness modelling, as well as from the
presence of gradients in population content. One physical mechanism
that can give rise to these gradients is the inside-out/outside-in
quenching.  For instance, inside-out quenching will result in younger
populations outside of the reach of the fibre \citep{2018GAMAGV2}.
The opposite will be true when galaxy is undergoing outside-in
quenching. Another cause for such behaviour would be galaxy
morphology. More specifically, spiral galaxies with a strong bulge
will produce a greater difference between fibre and whole galaxy
estimates of both the colour and D$_n$(4000).

\section{SDSS Green Valley Probability Catalogue}
\label{app:GV_Prob}

Adopting the methodology laid out in A19, we are able to
assign to each galaxy, on the D$_n$(4000) vs velocity dispersion plane,
a probability of it belonging to the Blue Cloud ($\hat{ \cal P}_{BC}$),
Green Valley ($\hat{ \cal P}_{GV}$), or Red Sequence 
($\hat{ \cal P}_{RS}$). Each probability can be written:
\begin{equation}
  \hat{\cal P}_k = \eta_k {\cal P}_k(\pi;\sigma),
  \label{eq:Fin_Prob}
\end{equation}
where ${\cal P}_k$ is given by Eq.~\ref{eq:Prob_Gen} and $\eta_k$ is a weight factor 
dependent on the number of galaxies in BC, GV or RS.

To assign a probability to an individual galaxy, we firstly obtain
the PDF averages: $\mu_{BC}$, $\mu_{GV}$, $\mu_{RS}$, and ``widths'':
$s_{BC}$, $s_{GV}$ and $s_{RS}$ for 11 bins from 50 to 350\,km/s. 
Then we interpolate between the velocity dispersion bins with 
 a third order polynomial that enables 
us to find the effective $\mu_k$ and $s_k$ in each group, BC, GV and RS, at 
any velocity dispersion. For each galaxy, we obtain its ${\cal P}_k({\cal G}_i)$,
along with $\mu_k({\cal G}_i)$ and $s_k({\cal G}_i)$. To determine $\eta_k$, we select all
galaxies with velocity dispersion within an interval centered at the
$\sigma$ of the chosen galaxy, with width $\Delta\sigma$=15\,km/s, and
then assign them either to the BC, GV
or RS, following the methodology described in A19, and in Sec.~\ref{ssec:GVDef} of this paper.
For any galaxy sitting on adjacent regions, 
we assign them evenly to their overlapping regions; e.g., if they overlap
between BC and GV, they are equally likely to count towards the BC or GV.
Thus we obtain $\eta_k$ for each galaxy. Once we obtain $\hat{\cal P}_k$, we normalise
the total probability to unity.
Note we assign galaxies with D$_n$(4000)$>$1.99,  $\hat{\cal P}_{RS}$=1, 
and $\hat{\cal P}_{BC} = \hat{\cal P}_{GV}$=0, as we know that galaxies with a high
4000 \AA\ break strength are old and quiescent. Likewise galaxies with D$_n$(4000)$<$0.8
are assigned $\hat{\cal P}_{BC}$=1, and $\hat{\cal P}_{GV}=\hat{\cal P}_{RS}$=0. 

\begin{figure}
  \centering
  \includegraphics[width=0.9\linewidth]{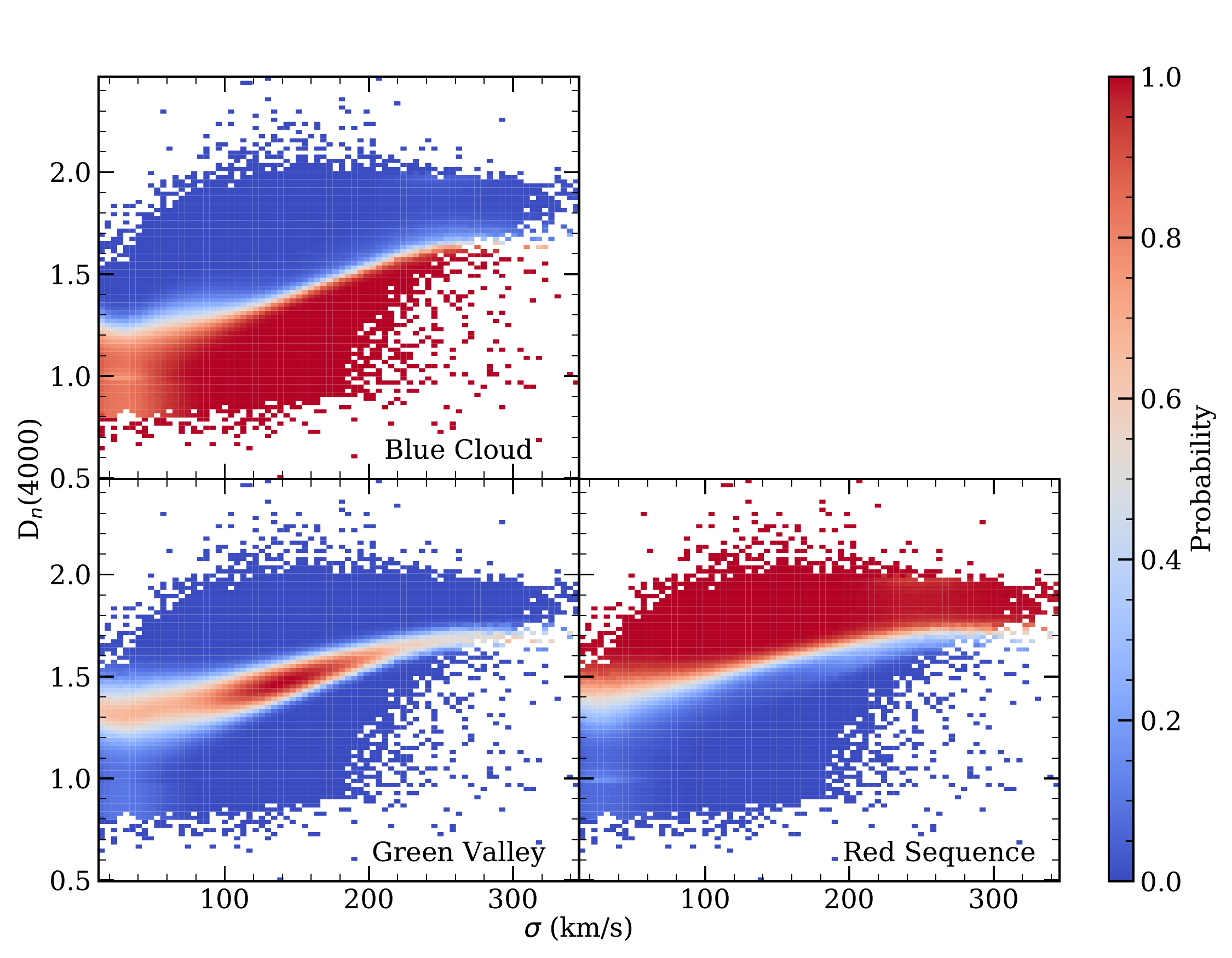}
  \caption{The probability of galaxies belonging to the Blue Cloud
    ({\it top left}), Green Valley ({\it bottom left}) and Red
    Sequence ({\it bottom right}). Blue colours represent galaxies
    with a low probability of residing in the three regions, as
    labelled, while red shows high probability (see colour bar for
    reference). For ease of visualisation, the resulting density plots
    have been modified with a Gaussian smoothing kernel.}
  \label{fig:Probs}
\end{figure}

A small number of galaxies from the catalogue is shown in Tab. \ref{tab:Probs} for reference.
The full catalogue can be found in the online version of the paper. 
Note owing to the methology used, the probabilites presented here are not 
unique solutions, but a realisation of the three groups. Thus if
one were to carry out the outlined steps, the results should not be identical, but
statistically equivalent. 
Fig.~\ref{fig:Probs} illustrates the probability density of galaxies residing in BC, GV or RS,
as presented in A19 and this paper.

\begin{table*}
  \centering
  \caption{Probabilities of BC, GV or RS membership
      assigned to each galaxy from the SDSS set. This table shows
      a small portion of the full set, giving the SDSS spectral
      identification (plate, Julian date and fibre ID) along with the
      4000\AA\ break strength and velocity dispersion
      (measured within the 3\,arcsec diameter fibres of the spectrograph).
      The final three columns list the normalized probability of belonging
      to either Blue Cloud, Green Valley, or Red Sequence, respectively,
      using the methodology presented here and in A19.}
  \label{tab:Probs}
  \begin{tabular}{rrrccrrr}
    \hline
    \multicolumn{3}{c}{SDSS ID} &  & Velocity & \multicolumn{3}{c}{Probabilities}\\
    plate & mjd & fibre ID & D$_n$(4000) & dispersion &$\hat{\cal P}_{BC}$ &$\hat{\cal P}_{GV}$ & $\hat{\cal P}_{RS}$\\
    & & & & km/s & & &\\
    \hline
    315 & 51663 & 560 & 1.012 &  64.5 & 0.970 & 0.009 & 0.021\\
    326 & 52375 &  86 & 1.584 &  99.2 & 0.001 & 0.026 & 0.973\\
    326 & 52375 & 119 & 1.509 & 104.6 & 0.011 & 0.360 & 0.629\\
    326 & 52375 & 150 & 1.777 & 168.2 & 0.000 & 0.000 & 1.000\\
    327 & 52294 & 147 & 1.651 & 147.9 & 0.001 & 0.013 & 0.986\\
    \hline
  \end{tabular}
\end{table*}

\bsp	
\label{lastpage}
\end{document}